\documentclass[preprints,article,accept,LaTeX,dvi2pdf]{Definitions/mdpi} 
\firstpage{1} 
\pubvolume{1}
\issuenum{1}
\articlenumber{0}
\pubyear{2022}
\copyrightyear{2022}
\datereceived{} 
\dateaccepted{} 
\datepublished{} 
\hreflink{https://doi.org/} 

\usepackage{subfig}
\usepackage{stmaryrd}
\usepackage[ruled]{algorithm2e} 
\usepackage{tensor}
\usepackage{float}
\usepackage{tikz-cd}

\newcommand{\R}{\mathbb{R}}

\newcommand{\E}{\mathbb{E}}
\newcommand{\Pb}{\mathbb{P}}
\newcommand{\Qb}{\mathbb{Q}}

\newcommand{\cF}{\mathcal{F}}

\newcommand{\cG}{\mathfrak{g}}

\newcommand{\cL}{\mathcal{L}}

\newcommand{\cFM}{\mathcal{FM}}

\newcommand{\partiel}[1]{\frac{\partial}{\partial {#1}}}

\DeclareMathOperator{\SPD}{SPD}
\DeclareMathOperator{\Log}{Log}
\DeclareMathOperator{\Exp}{Exp}
\DeclareMathOperator{\Vol}{Vol}
\DeclareMathOperator{\Cut}{Cut}

\DeclareMathOperator{\grad}{grad}
\DeclareMathOperator{\divergence}{div}

\usepackage{mathtools}
\DeclarePairedDelimiterX{\norm}[1]{\lVert}{\rVert}{#1}

\newcommand{\refsec}[1]{Section~\ref{#1}}

\DeclareMathOperator{\ad}{\mathrm{ad}}
\DeclareMathOperator{\Ad}{\mathrm{Ad}}

\newcommand{\GL}{\mathrm{GL}}

\newcommand{\On}[1]{\mathrm{O}({#1})}
\newcommand{\SO}{\mathrm{SO}}

\newcommand{\Symp}{\mathrm{Sym^+}}

\Title{Bridge simulation and metric estimation on Lie groups and homogeneous spaces}

\TitleCitation{Bridge simulation and metric estimation on Lie groups and homogeneous spaces}

\newcommand{\orcidauthorB}{0000-0001-6784-0328} 

\Author{Mathias H\o jgaard Jensen $^{1,\dagger,\ddagger}$, Lennard Hilgendorf $^{1,\dagger,\ddagger}$, Sarang Joshi $^{2,\ddagger}$, and Stefan Sommer $^{1,\ddagger}$\orcidauthorB}

\AuthorNames{Mathias H\o jgaard Jensen and Stefan Sommer}

\AuthorCitation{Jensen et al.}

\address{%
$^{1}$ \quad Department of Computer Science, University of Copenhagen; \{matje,lenhil,sommer\}@di.ku.dk
\\
$^{2}$ \quad Department of Biomedical Engineering, University of Utah; sjoshi@sci.utah.edu}

\corres{Correspondence: sommer@di.ku.dk}

\firstnote{Current address: Universitetsparken 5, DK-2100, Copenhagen E, Denmark} 
\secondnote{These authors contributed equally to this work.}

\abstract{We present schemes for simulating Brownian bridges on complete and connected Lie groups and homogeneous spaces. We use this to construct an estimation scheme for recovering an unknown left- or right-invariant Riemannian metric on the Lie group from samples. We subsequently show how pushing forward the distributions generated by Brownian motions on the group results in distributions on homogeneous spaces that exhibit non-trivial covariance structure. The pushforward measure gives rise to new parametric families of distributions on commonly occurring spaces such as spheres and symmetric positive tensors. We extend the estimation scheme to fit these distributions to homogeneous space-valued data. We demonstrate both the simulation schemes and estimation procedures on Lie groups and homogenous spaces, including $\SPD(3) = \GL_+(3)/\SO(3)$ and $\mathbb S^2 = \SO(3)/\SO(2)$.
}


\begin{document}

\section{Introduction}
Bridge simulation is a data augmentation technique for generating missing trajectories of continuous diffusion processes. We consider bridge simulation on Lie groups and homogeneous spaces. As an important example, we investigate the case of i.i.d. Lie group or homogeneous space valued samples that are considered discrete-time observations of a continuous diffusion process. Assuming the stochastic dynamics to be Brownian motion, we wish to estimate the underlying Riemannian metric of the Lie group or homogeneous space from the samples. To evaluate and maximize the likelihood of the data, we need to account for the diffusion process being unobserved at most time points. This requires bridge sampling, and the sampling techniques are thus the key enabler for metric estimation in this setting.

Simulation of conditioned diffusion processes is a highly non-trivial problem, even in Euclidean spaces. Because transition densities of diffusion process are only available in closed form for a narrow range of processes, simulating directly from the true bridge distribution is generally infeasible. 
The data augmentation used in inference for diffusion processes dates back to the seminal paper by Pedersen \cite{pedersen1995consistency} almost three decades ago. Since then, several papers have described diffusion bridge simulation methods; see, e.g., \cite{bladt2014simple, bladt2016simulation, bui2021inference, delyon_simulation_2006, jensen2019simulation, jensen_simulation_2021, van2017bayesian, papaspiliopoulos2012importance, schauer2017guided, sommer_bridge_2017}. The method by Delyon and Hu \cite{delyon_simulation_2006} exchanged the intractable drift term in the conditioned diffusion with a tractable drift originating from the drift of a Brownian bridge. Several papers have built on the ideas of Delyon and Hu. In particular, a manifold equivalent drift term analogous to the drift term of a Brownian bridge in Euclidean space was used in the paper \cite{jensen2019simulation} to describe the simulation of Brownian bridges on the flat torus, while \cite{jensen_simulation_2021} generalized this method to Riemannian manifolds. \cite{sommer_bridge_2017} used the drift to model Brownian bridges on the space of landmarks, and Bui et al. \cite{bui2021inference} used a similar drift term on the space of symmetric positive definite (covariance) matrices. The authors of the latter paper exploits the exponential map, which in the space of covariance matrices is a global diffeomorphism avoiding the cumbersome geometric local time. The details are further described in \cite{bui2022inference}.  

The idea of the present paper is based on the method presented in the paper by Delyon and Hu \cite{delyon_simulation_2006}, and, in the geometric setting, the paper \cite{jensen_simulation_2021}. When conditioning a diffusion with transition density $p_t$ to hit a point $v$ at time $T>0$, the intractable guiding drift term $\nabla_{x = X_t} \log p_{T-t}(x,v)$ in the stochastic differential equation (SDE) for the conditioned diffusion can be exchanged with the guiding drift term in the SDE for a Brownian bridge. This paper extends this idea to Lie groups and homogeneous spaces. 

\begin{figure}[t]
  \subfloat{
        \includegraphics[width=.23\linewidth,clip=true,trim=150 150 150 150]{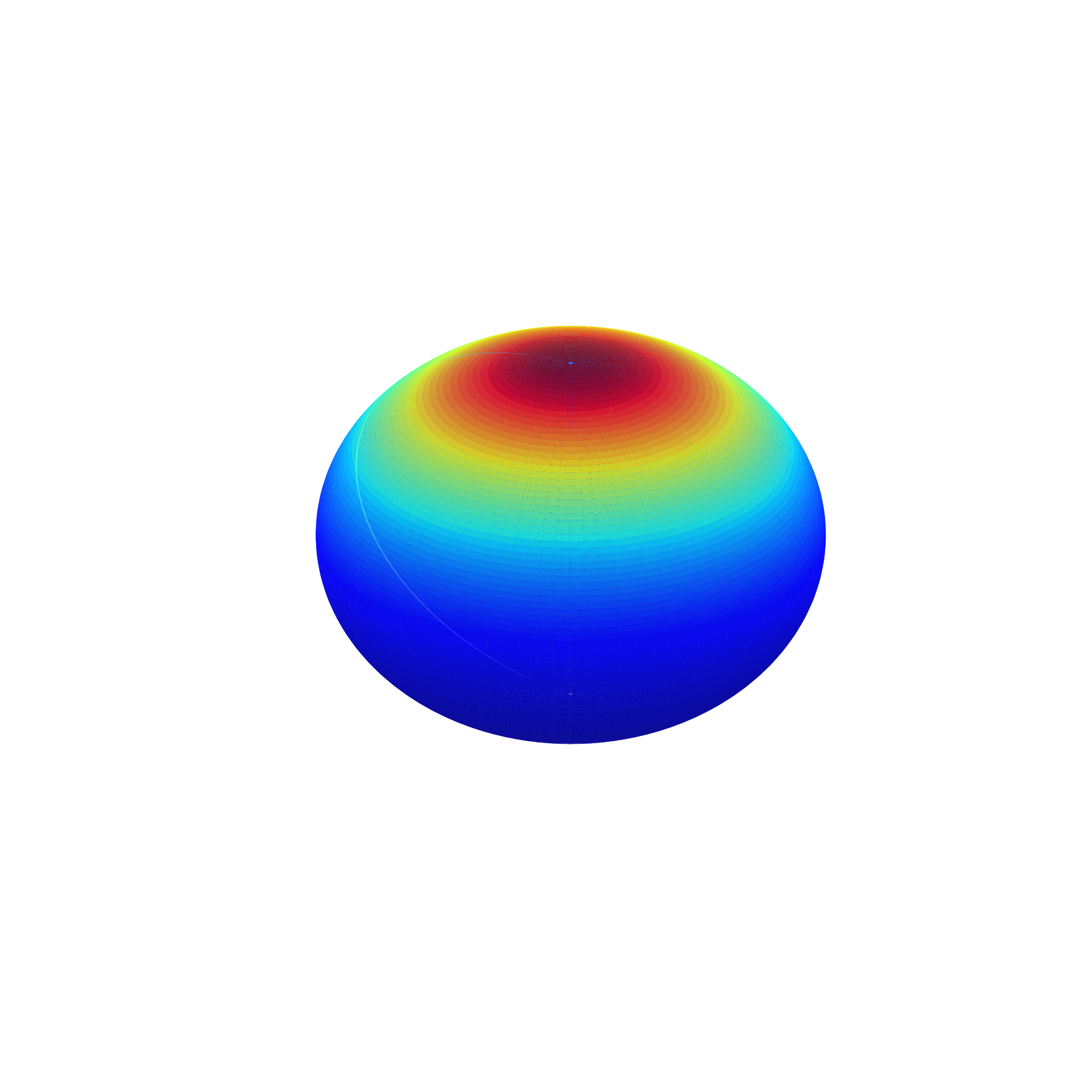}
      }
    \hfill
    \subfloat{
    \includegraphics[width=.23\linewidth,clip=true,trim= 150 150 150 150]{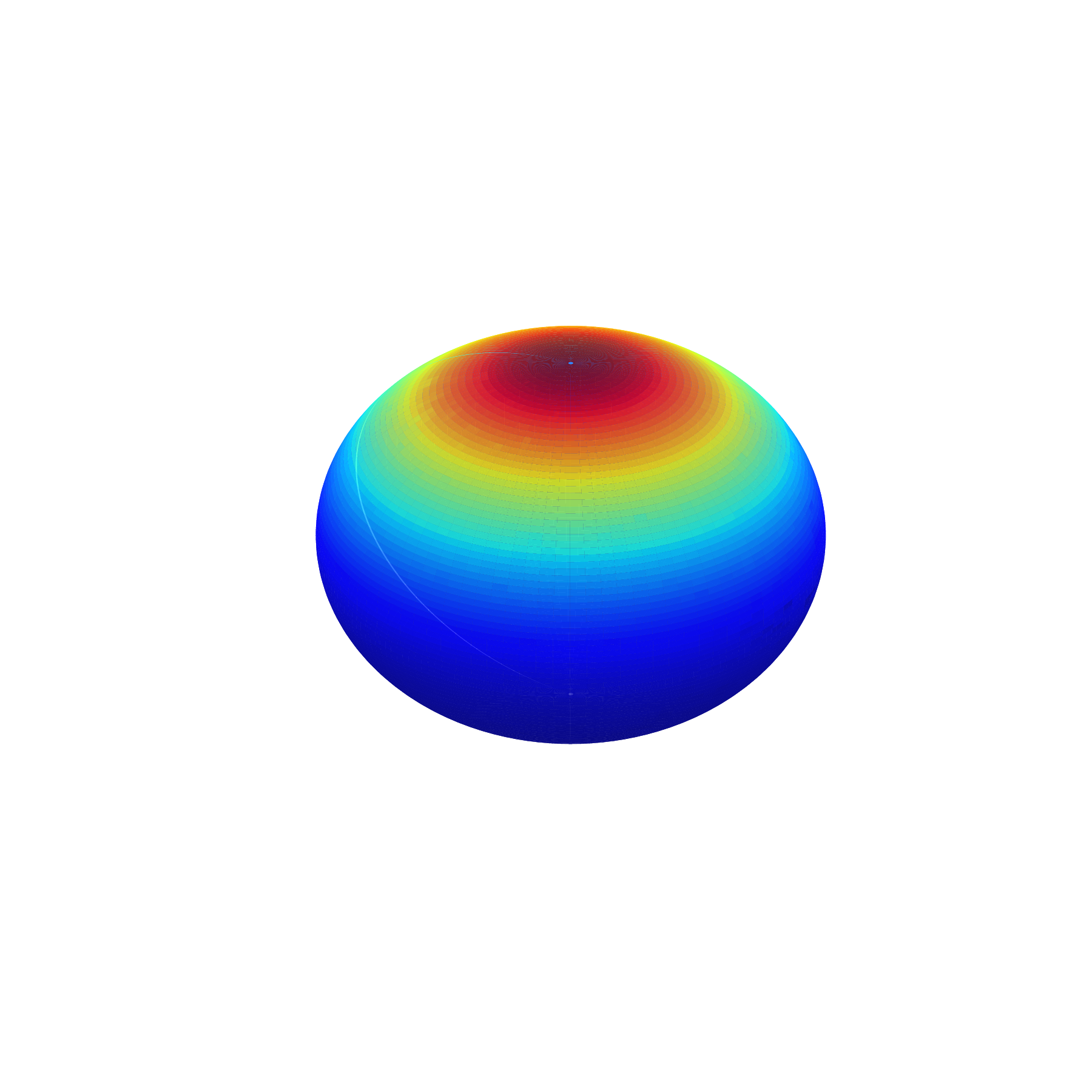}
  }
  \hfill
    \subfloat{
    \includegraphics[width=.23\linewidth,clip=true,trim= 150 150 150 150]{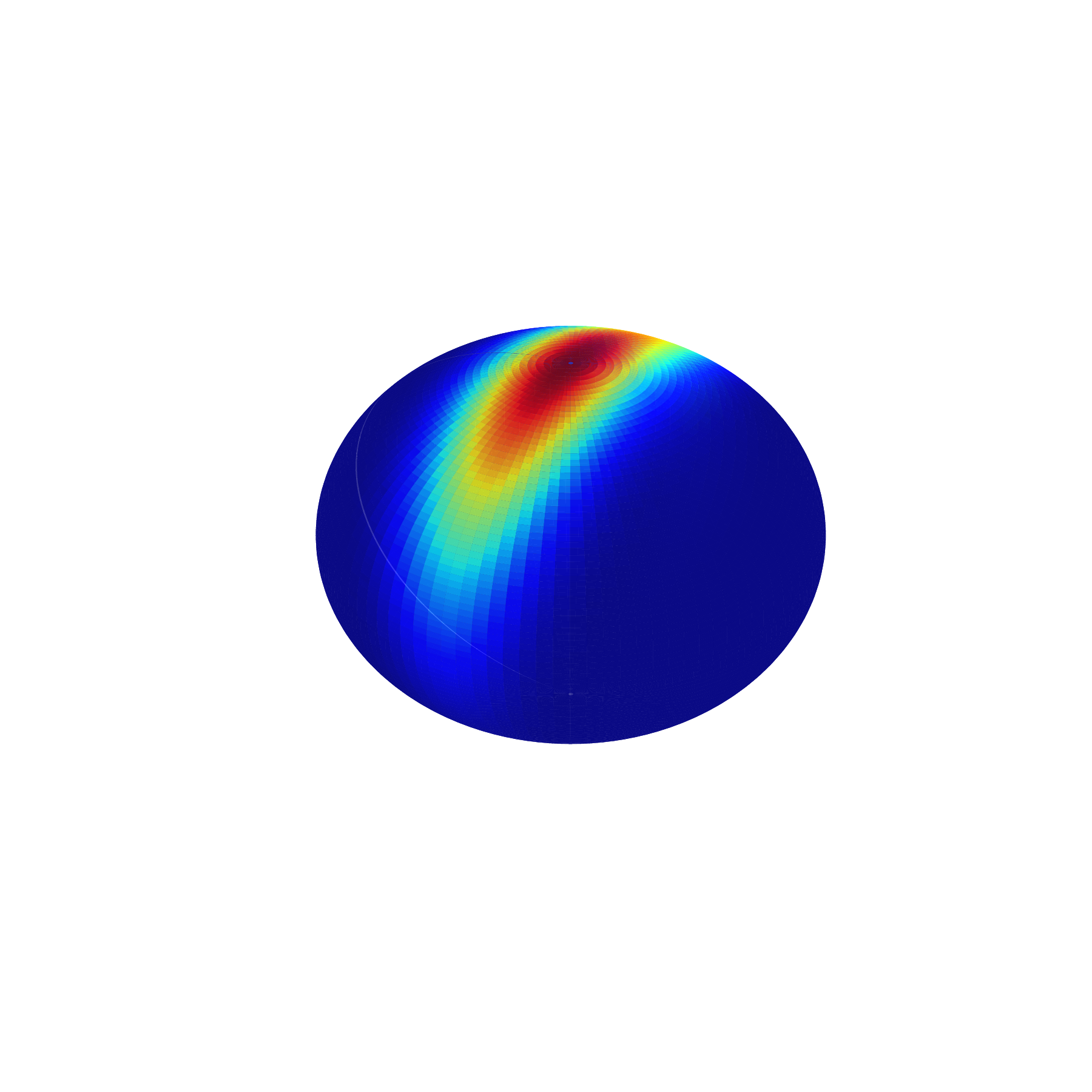}
  }
  \hfill
    \subfloat{
    \includegraphics[width=.23\linewidth,clip=true,trim= 150 150 150 150]{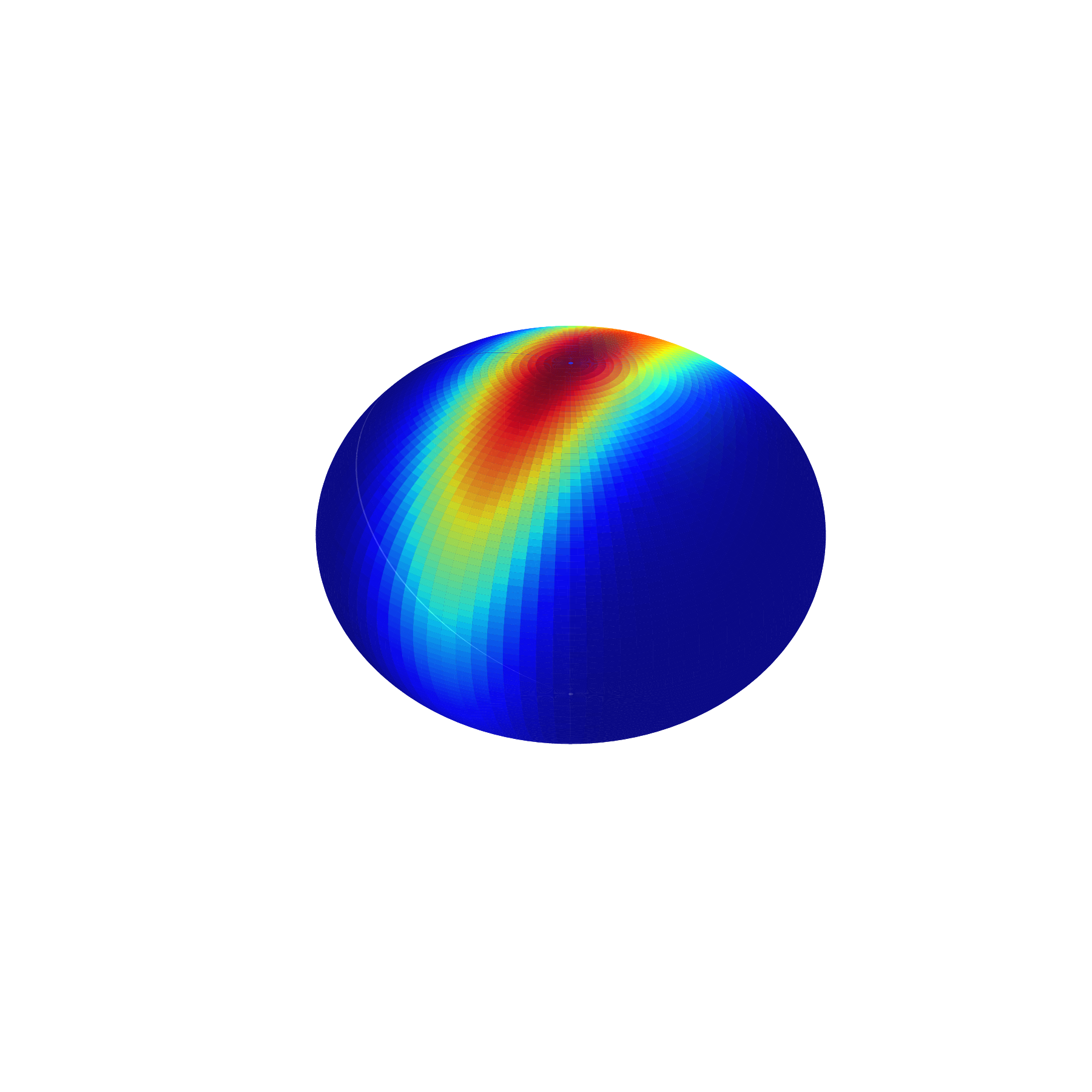}
  }
    \caption{The two leftmost plots visualize the transition densities of a Fisher-Bingham-Kent distribution (left) and the pushforward density of a Brownian motion to $\mathbb S^2$ (right) with a bi-invariant metric. The Fisher-Bingham-Kent distribution has been fitted to samples from a Brownian motion using MLE. The pushforward measure of a Brownian motion on $\SO(3)$ to the sphere $\mathbb S^2$ results in anisotropic distributions on $\mathbb S^2$ when the metric on $\SO(3)$ is not bi-invariant, here displayed in the two rightmost plots, for $T=0.5$ and $T = 1.0$, respectively.  
        The coloring indicate the density of the pushforward at different times (different color scheme for each subfigure).}
    \label{fig: anisotropic distribution on S2}
\end{figure}

As an application, we consider discrete-time observations in Lie groups and homogeneous spaces regarded as incomplete observations of sample paths of Brownian motions arising from left- (or right-) invariant Riemannian metrics. The bridge simulation schemes allow to interpolate between the discrete-time observations. Furthermore, we observe how varying the metric on Lie groups affects pushforwards of the Brownian motion to homogeneous spaces being quotients of the group. These distributions encode covariance of the data resulting from the metric structure of the Lie group. We define this family of distributions and derive estimation schemes for recovering the metric structure of the group both in the case of Lie group samples and in the case of homogeneous space samples.
One particular example is the two-sphere, $\mathbb S^2 \cong \SO(3)\backslash \SO(2)$. Changing the metric structure on $\SO(3)$ results in anisotropic distributions on $\mathbb S^2$, arising as the pushforward measure from $\SO(3)$. Figure \ref{fig: anisotropic distribution on S2} illustrates the isotropic and anisotropic distributions on $\mathbb S^2$ induced by a bi-invariant and left-invariant (non-invariant) metric on $\SO(3)$, respectively. The resulting distributions are analogous to the Von Mises-Fisher and Fisher-Bingham distributions \cite{fisher_dispersion_1953,kent_fisher-bingham_1982}. However, unlike the Von Mises-Fisher and Fisher-Bingham distributions, the approach is independent of the chosen embedding instead resulting from the quotient structure. 

For simulation on homogeneous spaces, we present three schemes. The first builds on the idea of Thompson \cite{thompson2018brownian} by conditioning on a submanifold in the Lie group $G$ obtained as a fiber over the point $v \in M = G/K$, for some closed subgroup $K \subseteq G$. The second scheme assumes the homogeneous space has a discrete fiber $\Gamma$ and therefore, the fiber over $v \in M$, $\pi^{-1}(v)$, is discrete. Using the $k$-nearest-points from the fiber $\pi^{-1}(v)$ to the initial point $x_0$, we obtain a truncated guiding drift term convergence to a subset of $\pi^{-1}(v)$. The last scheme assumes that the fiber is connected. By sampling a finite number of points in the fiber over $v$, a similar conditioning is obtained.



Statistics on Lie groups and homogeneous spaces finds applications in many diverse fields including bioinformatics, medical imaging, shape analysis, computer vision, and information geometry, see, e.g., \cite{garcia-portugues_langevin_2017, hamelryck_sampling_2006, pennec_riemannian_2006,vaillant_statistics_2004, yang_means_2011}. Statistics in Euclidean spaces often relies on the distributional properties of the normal distribution. Here we use Brownian motions and the heat equation to generalize the normal distribution to Lie groups and homogeneous spaces as introduced by Grenander \cite{grenander_probabilities_1963}. The solution to the heat equation is the transition density of a Brownian motion. Through Monte Carlo simulations of bridges, we can estimate the transition density and maximize the likelihood with respect to the Riemannian metric.



\subsection{Contribution and Overview}
We present simulation schemes on Lie groups and homogeneous spaces with application to parameter estimation. We outline the necessary theoretical background for the construction of bridge simulation on Lie groups and homogeneous spaces before demonstrating how the simulation scheme leads to estimates of means and underlying metric structure using maximum likelihood estimation on certain Lie groups and homogeneous spaces. 
The paper builds on and significantly extends the conference paper \cite{jensen_simulation_2021} that introduced bridge simulation in the Lie group setting. 

The paper is organized as follows. In \refsec{sec: Notation and Background}, we describe the relevant background theory of Lie groups, Brownian motions, and Brownian bridges in Riemannian manifolds. \refsec{sec: simulation of bridges on Lie groups} presents the theory and results of bridge sampling in Lie groups, while \refsec{sec:simulation of bridges on Lie groups and Homogeneous spaces} introduces bridge sampling on various homogeneous spaces.
Numerical experiments on certain Lie groups and homogeneous spaces are presented in \ref{sec: numerical experiments}. 
\section{Notation and background}\label{sec: Notation and Background}
We here briefly describe simulating conditioned diffusions in $\R^n$ as developed in \cite{delyon_simulation_2006} before reviewing the theory on conditioned diffusion on Riemannian manifolds.

\subsection{Euclidean diffusion bridges and simulation}
Suppose  a strong solution to an SDE of the form
\begin{align*}
    dx_t = b(t,x_t)dt + \sigma(t,x_t) dw_t,
\end{align*}
where $b$ and $\sigma$ satisfies certain regularity conditions and where $w$ denote a $\R^n$-valued Brownian motion. In this case, $x$ is Markov, and its transition density exists. Suppose we define the function    
    \begin{align*}
        h(t,x) = \frac{p_{T-t}(x_t,v)}{p_T(x_0,v)},
    \end{align*}
for some $x_0, v \in \R^n$. Then it is easily derived that $h$ is a martingale on $[0,T)$ with $h(0,x_0)=1$ and Doob's $h$-transform implies that the SDE of the conditioned diffusion $x|x_T=v$ is given by
    \begin{align*}
        dy_t = \tilde{b}(t,y_t) dt + \sigma(t,y_t)dw_t
    \end{align*}
where $\tilde{b}(t,y) = b(t,y) + (\sigma \sigma^T)(t,y) \nabla_y \log p_{T-t}(y,v)$. In case that the transition density is intractable, simulation from the exact distribution is in-feasible. Delyon and Hu \cite{delyon_simulation_2006} suggested substituting the latter term in $\tilde b$ with a drift term of the form $-(y_t-v)/(T-t)$, which equals the drift term in a Brownian bridge. The guided process obtained by making the above substitution yields a conditioning and one obtain 
\begin{equation}
    \E[f(x)|x_T=v] = C\E[f(y) \varphi_T(y)],
\end{equation}
where $\varphi_T$ is a likelihood function that is tractable and easy to compute, $y$ is the guided process, and the constant $C > 0$ depends on $x_0$, $v$, and $T$. 

\subsection{Riemannian manifolds and Lie groups}


Let $M$ be a finite dimensional smooth manifold of dimension $d$. 
$M$ can be endowed with a Riemannian metric tensor, i.e., a family of inner products $\{\langle \cdot, \cdot \rangle_x\}_{x\in M}$ defined on each tangent space $T_xM$. The Riemannian metric tensor gives rise to a distance function between points in $M$. The tangent space is locally diffeomorphic with an open subset of $M$. The Riemannian exponential map $\Exp_x \colon T_xM \rightarrow M$ provides this local diffeomorphism. On the subset of $M$ where $\Exp_x$ is a diffeomorphism the inverse Riemannian exponential map, also called the Riemannian logarithm map, $\Log_x \colon M \rightarrow T_xM$ is defined. The Riemannian distance function can then be defined in terms of the Riemannian inner product as $d(x,y) = \lVert \Log_x(y)\rVert_x$. The Riemannian logarithm map plays an important role when defining guided bridges on manifolds. 

Let $X$ be a vector field on $M$ assigning to each point $X \in M$ a tangent vector $X(x)\in T_xM$. A connection $\nabla$ on a manifold is an operation that allows us to compare neighboring tangent spaces and define derivatives of vector fields along other vector fields, that is, if $Y$ is another vector field, then $\nabla_X Y$ is the derivative of $Y$ along $X$ (also known as the covariant derivative of $Y$ along $X$). A connection also gives a notion of "straight lines" in manifolds, also known as geodesics. A curve $\gamma$ is a geodesic if the vector field along $\gamma$ is parallel to itself, i.e, if $\nabla_{\dot \gamma_t} \dot \gamma_t = 0$. The geodesic curves are locally length minimizing. 

Generalizing the Euclidean Laplacian operator, the Laplace-Beltrami operator is defined as the  divergence of the gradient, $\Delta_M f=\divergence\grad f$. In terms of local coordinates $(x_1,\dots,x_d)$ the expression for the Laplace-Beltrami operator becomes
\begin{align}\label{laplace beltrami}
\Delta_M f = \det(g)^{-1/2}\left(\frac{\partial}{\partial x_j} g^{ji} \det(g)^{1/2} \frac{\partial}{\partial x_i}\right)f,
\end{align}
where $\det(g)$ denotes the determinant of the Riemannian metric $g$ and $g^{ij}$ are the coefficients of the inverse of $g$. \eqref{laplace beltrami} can be written as
\begin{align}\label{laplace beltrami in diffusion operator form}
\Delta_Mf &= a^{ij}\frac{\partial}{\partial x_i} \frac{\partial}{\partial x_j} f + b^j \frac{\partial}{\partial x_j} f,
\end{align}
where $a^{ij}=g^{ij}$, $b^k = - g^{ij}\Gamma^k_{ij}$, and $\Gamma$ denote the Christoffel symbols of the Riemannian metric. 

\subsection{Lie groups}
Let $G$ denote a connected Lie group of dimension $d$, i.e., a smooth manifold with a group structure such that the group operations $G \times G \ni (x,y) \overset{\mu}{\mapsto} xy \in G$ and $G \ni x \overset{\iota}{\mapsto} x^{-1}\in G$ are smooth maps. If $x \in G$, the left-multiplication map, $L_x y$, defined by $y \mapsto \mu(x,y)$, is a diffeomorphism from $G$ to itself. Similarly, the right-multiplication map $R_x y$ defines a diffeomorphism from $G$ to itself by $y \mapsto \mu(y,x)$. 
Let $dL_x \colon TG \rightarrow TG$ denote the pushforward map given by $(dL_x)_y \colon T_yG \rightarrow T_{xy}G$. A vector field $V$ on $G$ is said to be left-invariant if $(dL_x)_yV(y) = V(xy)$. The space of left-invariant vector fields is linearly isomorphic to $T_eG$, the tangent space at the identity element $e \in G$. By equipping the tangent space $T_eG$ with the Lie bracket we can identify the Lie algebra $\mathfrak g$ with $T_eG$. The group structure of $G$ makes it possible to define an action of $G$ on its Lie algebra $\mathfrak g$. The conjugation map $C_x := L_x \circ R_x^{-1} \colon y \mapsto xyx^{-1}$, for $x \in G$, fixes the identity $e$. Its pushforward map at $e$, $(dC_x)_e$, is then a linear automorphism of $\mathfrak g$. Define $\Ad(x) := (dC_x)_e$, then $\Ad \colon x \mapsto \Ad(x)$ is the adjoint representation of $G$ in $\cG$. The map $G \times \cG \ni (x,v) \mapsto \Ad(x)v \in \mathfrak g$ is the adjoint action of $G$ on $\mathfrak g$. We denote by $\langle \cdot , \cdot \rangle$ a Riemannian metric on $G$. The metric is said to be left-invariant if $\langle u, v \rangle_y = \left\langle (dL_x)_y u, (dL_x)_y v \right\rangle_{L_x(y)}$, for every $u,v \in T_yG$, i.e., the left-multiplication maps are isometries, for every $x \in G$. The metric is $\Ad(G)$-invariant if  $\langle u, v \rangle_e = \left\langle \Ad(x) u, \Ad(x) v \right\rangle_e$, for every $u,v \in \mathfrak g$. Note that an $\Ad(G)$-invariant metric on $G$ is equivalent to a bi-invariant (left- and right-invariant) inner product on $\mathfrak g$. The differential of the $\Ad$ map at the identity yields a linear map $\ad(x) = \frac{d}{dt} \Ad(\exp(tx))|_0$. This linear map is equal to the Lie bracket $[v,w] = \ad(v)w$, $v,w \in \cG$.

A one-parameter subgroup of $G$ is a continuous homomorphism $\gamma \colon (\R,+) \rightarrow G$. The Lie group exponential map $\exp \colon \mathfrak g \rightarrow G$ is defined as $\exp(v) = \gamma_v(1)$, for $v \in \cG$, where $\gamma_v$ is the unique one-parameter subgroup of $G$ whose tangent vector at $e$ is $v$. For matrix Lie groups the exponential map has the particular form: $\exp(A) = \sum_{k=0}^\infty A^k/k!$, for a square matrix $A$. The resulting matrix $\exp(A)$ is an invertible matrix. Given an invertible matrix $B$, if there exists a square matrix $A$ such that $B = \exp(A)$, then $A$ is said to be the logarithm of $B$. In general, the logarithm might not exist and if it does it may fail to be unique. However, the matrix exponential and logarithms can be computed numerically efficient (see \cite[Chapter 5]{pennec_riemannian_2020} and references therein). 
In a neighborhood sufficiently close to the identity the Lie group logarithm exist and is unique. By means of left-translation (or right-translation), the Lie group exponential map can be extended to a map $\exp_g \colon T_gG \rightarrow G$, for all $g \in G$, defined by $\exp_g(v) = g \exp(dL_{g^{-1}}v)$. Similarly, the Lie group logarithm at $g$ becomes $\log_g(v) = dL_g \log(g^{-1}v)$.

\begin{Example}
    A few examples of Lie groups include the Euclidean space $(\R^n,+)$ with the additive group structure, $(\R_+, \cdot)$ the positive real line with a multiplicative group structure, the space of invertible real matrices $\GL(n)$ equipped with a multiplication of matrices forms a Lie group, and the rotation group $\On{n}$, consisting of real orthogonal matrices with determinant one or minus 1 forms a subgroup of $\GL(n)$.
\end{Example}

The identification of the space of left-invariant vector fields with the Lie algebra $\cG$ allows for a global description of $\Delta_G$. Indeed, let $\{v_1,\dots v_d\}$ be an orthonormal basis of $T_eG$. Then $V_i(g) = (dL_g)_e v_i$ defines left-invariant vector fields on $G$ and the Laplace-Beltrami operator can be written as (cf. \cite[Proposition 2.5]{liao2004levy})
\begin{equation*}
    \Delta_G f(e) = \sum_{i=1}^d V_i^2f(e) - V_0 f(e),
\end{equation*}
where $V_0 = \sum_{i,j=1}^d C_{ij}^j V_j$ and $C^k_{ij}$ denote the structure coefficients given by
\begin{equation}
    [V_i,V_j] = C^k_{ij}V_k.
\end{equation}
By the left-invariance, the formula for the Laplace-Beltrami operator holds globally, i.e., $\Delta_G f(g) = \Delta_G f \circ L_g(e) = \left(dL_g\right)_e \Delta_G f(e)$.

\subsection{Homogeneous spaces}
A homogeneous space is a particular type of quotient manifold that arises as a smooth manifold endowed with a transitive smooth action by a Lie group $G$. The homogeneous space is called a $G$-homogeneous space to indicate the Lie group action. All $G$-homogeneous spaces arise as a quotient manifold $G/H$, for some closed subgroup $H \subseteq G$. $H$ is a closed subgroup of the Lie group $G$ which makes $H$ into a Lie group. Any homogeneous space is diffeomorphic to the quotient space $G/G_x$, where $G_x$ is the stabilizer for the point $x$. The dimension of the $G$-homogeneous space is equal to $\dim G - \dim H$ the quotient map $\pi \colon G \rightarrow G/H$ is a smooth submersion, i.e., the differential of $\pi$ is surjective at every point. This implies that the fibers $\pi^{-1}(x)$, $x\in M$ are embedded submanifolds of $G$. We assume throughout that $G$ acts on itself by left-multiplication. 

\begin{Example}
    The rotation group $\SO(n)$ acts transitively on $\mathbb S^{n-1}$, therefore $\mathbb S^{n-1}$ is a $\SO(n)$-homogeneous space. Consider a point in $\mathbb S^{-1}$ as a vector in $\mathbb R^n$. Rotations that fix the point occur precisely in the subspace orthogonal to the vector. Thus, the stabilizer or isotropy group is the rotation group $\SO(n-1)$ and $\mathbb S^{n-1} = \SO(n)/\SO(n-1)$. 
    
    The set of invertible matrices with positive determinant $\GL_+(n)$ acts on symmetric positive definite matrices $\SPD(n)$. The isotropy group is the rotation group $\SO(n)$ and thus $\SPD(n)=\GL_+(n)/\SO(n)$. 
    
    A particular type of homogeneous space arises when the subgroup is a discrete subgroup of $G$. 
    For example, the space $\mathbb T^n = \R^n/\mathbb Z^n$ defines the $n$-torus as a homogeneous space.
\end{Example}

\subsection{Brownian motion on Riemannian manifolds}\label{sec: brownian motion}

The Laplacian defines Brownian motion on $M$ as a $\frac12 \Delta_M$-diffusion process up to its explosion time $\tau$. 
The stochastic differential equation (SDE) for a Brownian motion $X_t$ in local coordinates is
    \begin{equation}\label{eq: brownian motion local}
        dX_t^k = -\frac{1}{2} g^{ij}(X_t)\Gamma_{ij}^k(X_t) dt  + \sigma^k_j(X_t)dB_t^j,
    \end{equation}
where $\sigma = \sqrt{g^{-1}}$ is the matrix square root of $g^{-1}$. 

On Lie groups, an SDE for a Brownian motion on $G$ in terms of left-invariant vector fields takes the form
    \begin{equation}\label{eq: BM SDE on Lie group}
        dg_t = - \frac{1}{2}V_0(g_t) dt + V_i(g_t) \circ dB^i_t, \qquad g_0 = e,
    \end{equation}
where $\circ$ denotes integration in the Stratonovich sense. By \cite[Proposition 2.6]{liao2004levy}, if the inner product is $\Ad(G)$ invariant, then $V_0 = 0$. The solution of \eqref{eq: BM SDE on Lie group} is conservative or non-explosive and is called the left-Brownian motion on $G$ (see \cite{shigekawa1984transformations} and references therein).

\subsection{Brownian bridges}

In this section, we briefly review some facts on Brownian bridges on Riemannian manifolds, including Lie groups. On Lie groups, the existence of left-invariant (resp. right-invariant) vector fields allows identification of the Lie algebra with the vector space of left-invariant vector fields making the Lie group parallelizable. This allows constructing general semimartingales directly on the Lie groups.

Let $\Pb^t_x = \Pb_x|_{\cF_t}$ be the measure of a Riemannian Brownian motion, $X_t$, at some time $t$ started at point $x$. Let $p_t$ denote the transition density of $X_t$ so that $d\Pb^t_x = p_t(x,y) d\Vol(y)$ with $d\Vol(y)$ the Riemannian volume measure. Conditioning the Riemannian Brownian motion to hit some point $v\in M$ at time $T>0$ defines a Riemannian Brownian bridge. We let $\Pb_{x,v}^T$ denote the corresponding probability measure. The two measures are absolutely continuous (equivalent) over the time interval $[0, T)$, however mutually singular at time $t=T$. This is an obvious consequence of the fact that $\Pb_x(X_T = v)=0$, whereas $\Pb_{x,v}^T(X_T = v) = 1$. The corresponding Radon-Nikodym derivative is 
	\begin{equation}\label{eq: radon-nikodym for v}
		\frac{d\Pb_{x,v}^T}{d\Pb_x}\big|_{\cF_s} 
		=
		\frac{p_{T-s}(X_s, v)}{p_T(x,v)} 
		\qquad 
		\text{for } 0 \leq s <T,
	\end{equation}
which is a martingale for $s < T$. The Radon-Nikodym derivative defines the density for the change of measure, and it provides the conditional expectation 
	\begin{equation}
		\E[F(X_t)|X_T = v] = \frac{\E[p_{T-t}(X_t,v)F(X_t)]}{p_T(x,v)},
	\end{equation}
for any bounded and $\cF_s$-measurable random variable $F(X_s)$. 
The Brownian bridge is a non-homogeneous diffusion on $M$ with infinitesimal generator 
	\begin{align*}
	\cL_s f(z) = \frac{t}{2} \Delta_M f(z) + t \nabla_z \log p_{t(1-s)}(z,v) \cdot \nabla f(z).
	\end{align*}
The bridge can be described by an SDE in the frame bundle $FM$ of $M$. Let $U_t$ be a lift of $X_t = \pi_{FM}(U_t)$ and, using the horizontal vector fields $H_i,\ldots,H_d$, we have 
	\begin{equation}\label{eq: BB sde on frame bundle}
	dU_t = H_i(U_t) \circ \left(dB^i_t + \left(U_t^{-1}\left(\pi_* \left(\nabla^H_{u|u=U_t} \log \tilde{p}_{T-t}(u,v)\right)\right)\right)^i dt \right), \quad U_0 = u_0, 
	\end{equation}
where $\tilde{p}_t(u,v) = p_t(\pi(u),v)$ denotes the lift of the transition density, $B$ is an $\R^d$-valued Brownian motion, and $(\pi_{FM})_* \colon T\cFM \rightarrow TM$ is the pushforward of the projection $\pi_{FM} \colon \cFM \rightarrow M$. Her $u_0\in FM$ is an orthonormal frame such that $\pi_{FM}(u_0)=x_0$.


It is possible to simulate from the conditioned process directly for specific homogeneous spaces. As an example, we mention the case of the flat torus $\mathbb T^2$ considered a homogeneous space of $\R^2$ with fibers the set of integers $\mathbb Z^2$. In this case, a Brownian motion in $\mathbb T^2$ conditioned at a point in $\mathbb T^2$ lifts to a bridge in $\R^2$ conditioned on a set of points isomorphic to $\mathbb Z^2$. 


A generalization of Riemannian Brownian bridges can be found in Thompson \cite{thompson2018brownian}. Brownian bridges to submanifolds are here introduced by considering the transition density on a Riemannian manifold $M$ defined by
	\begin{equation}\label{eq: density on submanifold}
	p_t(x,N):= \int_{N} p_t(x,y)d\Vol_{N}(y),
	\end{equation}
where $N \subset M$ is a submanifold of $M$ and $\Vol_{N}$ denotes the volume measure on $N$. 
These processes are denoted \textit{Fermi bridges}. They have infinitesimal generator 
\begin{equation}\label{eq: infinitesimal generator of fermi bridge}
    \tfrac{1}{2}\Delta - \tfrac{r_N}{T-t}\tfrac{\partial}{\partial r_N},
\end{equation}
where $r_N(\cdot) := d(\cdot,N) = \inf_{y\in N} d(\cdot,y)$ and $\tfrac{\partial}{\partial r_N} = \nabla d(\cdot,N)$. 
The resulting conditional expectation becomes
	\begin{equation}\label{eq: conditional expection on N}
	\E[f(X_t)|X_T \in N] = \frac{\E[p_{T-t}(X_t,N)f(X_t)]}{p_T(x,N)},
	\end{equation}
which holds for all bounded $\cF_t$-measurable random variables $f(X_t)$. \cite{jensen2022mean} exploited the above idea to estimate diffusion means on manifolds by conditioning on the diagonal of a product manifold. In the current paper, the fibers of homogeneous spaces are embedded submanifolds of a Lie group and a simulation scheme on homogeneous spaces are obtained by conditioning on the fibers.

\subsection{One-point motions}

Consider the homogeneous space $M = G/H$, where $H$ is a Lie subgroup of the Lie group $G$ and let $\pi \colon G \rightarrow M$ denote the canonical projection. Suppose that $G$ acts on $M$ on the left and that $g_t$ is a process in $G$. As described in Liao \cite{liao2004levy}, we obtain an induced process in $M$ induced by the process $g_t$ in $G$. For any $x \in M$, the induced process $x_t = g_tx$ defines the one-point motion of $g_t$ in $M$, with initial value $x$. 

The one-point motion, $X_t = g_tx$, of a Brownian motion $g_t$ in $G$, started at $g_0=e$, is only a Brownian motion in $M$ under certain regularity conditions (see \cite[Proposition 2.7]{liao2004levy}). In the case of a bi-invariant metric, a Brownian motion on $G$ maps to a Brownian motion in $M$ through its one-point motion. One-point processes might not even preserve the Markov property in the general case. 
	

\subsection{Pushforward measures}


Let $\pi \colon G \rightarrow M$ be the projection to the homogeneous space $M = G/H$. Then $\pi$ is a measurable map, and, if $\mu$ is a measure on $G$, the pushforward of $\mu$ by $\pi$, defined by $\pi_* \mu (B) = \mu\left(\pi^{-1}(B)\right)$, for all measurable subsets $B \subseteq M$, is a measure on $M$. 
A numerical example is provided in Figure~\ref{fig: anisotropic distribution on S2} showing anisotropic distributions on the homogeneous space $\mathbb S^2$ obtained from pushing forward Brownian motions of a non-invariant metric on the top space $\SO(3)$.

%
	
The Riemannian volume measure $\Vol_G$ on $G$ decomposes into a product measure consisting of the volume measure on fibers in $G$, e.g. $\pi^{-1}(z)$, and the volume measure on its horizontal complement, i.e., $d\Vol_G = d\Vol_{\pi^{-1}(z)} d\Vol_{|H}(z)$, where $d\Vol_{|H}$ is the horizontal restriction of the volume measure in $G$.
 The measure of a process $g_t$ on $G$ pushes forward to $M$, and we denote the corresponding density wrt. the volume measure on $M$ for $p_t^M$.
 Then $p_t^M(x)=\int_{\pi^{-1}(x)}p_t^G(g_0,y)d\Vol_{\pi^{-1}(z)}(y)$.

%
%
	
	\begin{Lemma}\label{lemma: conditional expectation on M}
	Let $g_t$ be a Markov process on $G$, started at $g_0 \in G$, with density $p^G_t(g_0,\cdot)$, and let $X_t=\pi(g_t)$. The conditional expectation on $M$ satisfies
	\begin{equation*}
	\E[f(X) | X_T = v] = \E\left[f( X) \frac{p^M_{T-t}( X_t, v)}{p_T^M( x_0,v)} \right],
	\end{equation*}
	for all bounded, continuous, and non-negative $\mathcal F_t$-measurable $f$ on $M$. Furthermore,
	    \begin{equation*}
	        \E[\tilde f(g) | g_T \in N] = \E[f(X) | X_T = v],
	    \end{equation*}
	 where $\tilde f = f \circ \pi$.
	\end{Lemma}

	\begin{proof}
	Let $f$ be a bounded, continuous, and non-negative measurable function on $M$, and let $\tilde f = f \circ \pi$. 
	Then it follows directly from \eqref{eq: radon-nikodym for v} and \eqref{eq: conditional expection on N} that 
	\begin{align*}
	\E[\tilde f(g)|g_T \in N ] 
	= &
	\E\left[f(\pi(g))\frac{p^G_{T-t}(g_t,N)}{p^G_T(g_0,N)}\right] 
	=
	\E\left[f(\pi(g_t))\frac{p^G_{T-t}(g_t,\pi^{-1}(v))}{p^G_T(g_0,\pi^{-1}(v))}\right] 
	\\
	=&
	\E\left[f(\pi(g_t))\frac{\pi_*p^G_{T-t}(g_t,v)}{\pi_* p^G_T(g_0,v)}\right] 
	=  
	\E\left[f(X_t)\frac{p^M_{T-t}(X_t,v)}{p^M_T( x_0,v)}\right].
	\end{align*}
	\end{proof}

\section{Simulation of bridges on Lie groups}\label{sec: simulation of bridges on Lie groups}

\begin{figure}[t]
\centering
        \includegraphics[width=.48\linewidth,clip=true,trim=250 250 200 250]{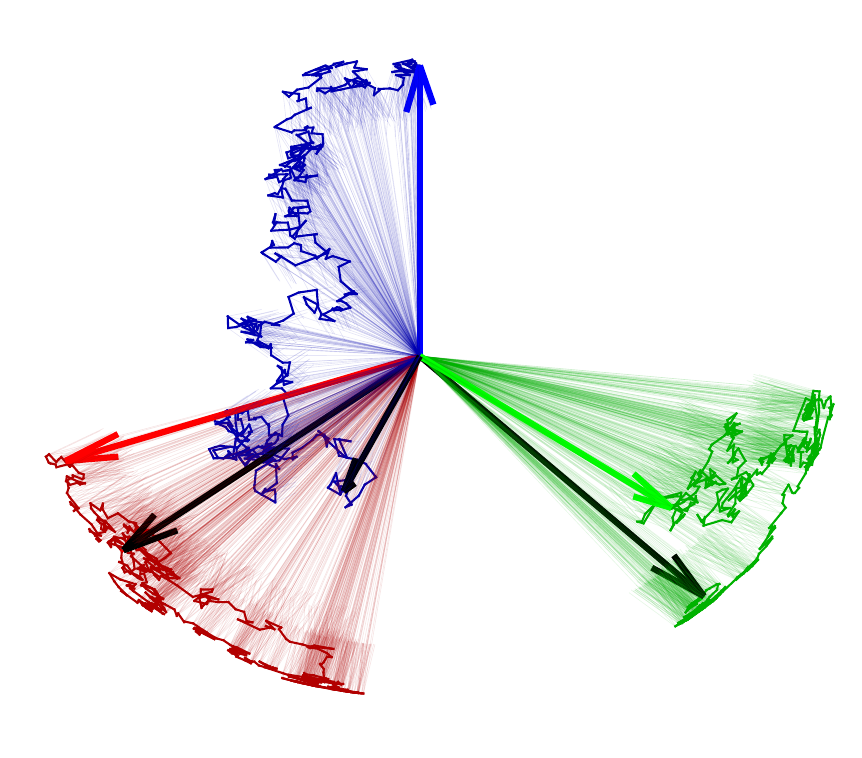}
    \caption{One sample path of the guided bridge process defined by \eqref{eq: guided diffusion sde} visualized by its action on basis vectors (red, blue, green) of $\R^3$. The bridge is conditioned on the rotation indicated by the black arrows. }
    \label{fig: bridge on so3}
\end{figure}

\begin{figure}[t]
  \subfloat{
        \includegraphics[width=.48\linewidth,clip=true,trim=250 250 200 250]{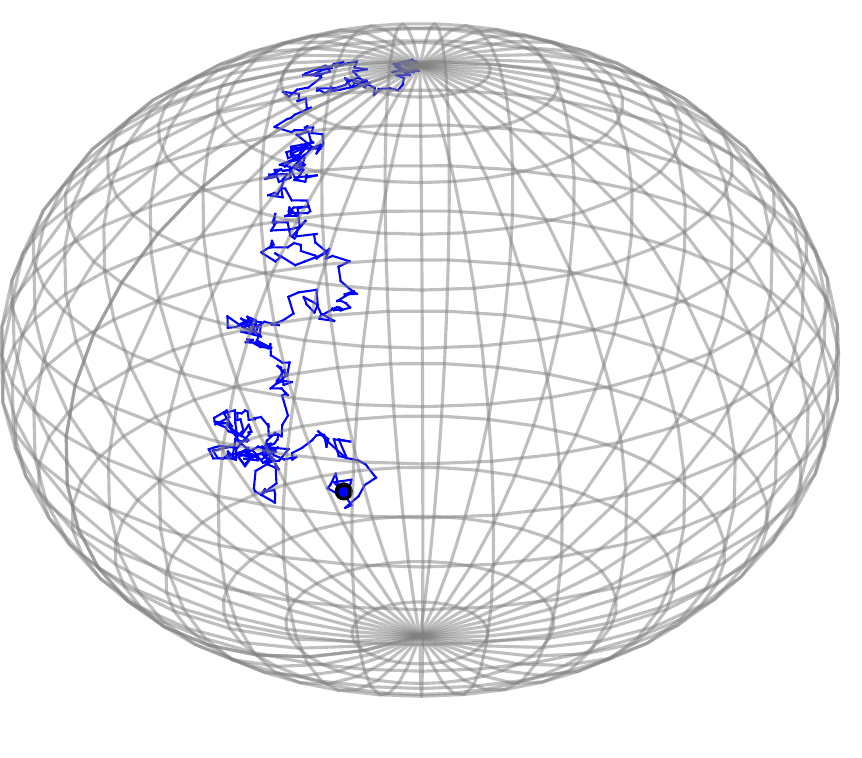}
      }
    \hfill
    \subfloat{
    \includegraphics[width=.48\linewidth,clip=true,trim= 75 75 75 100]{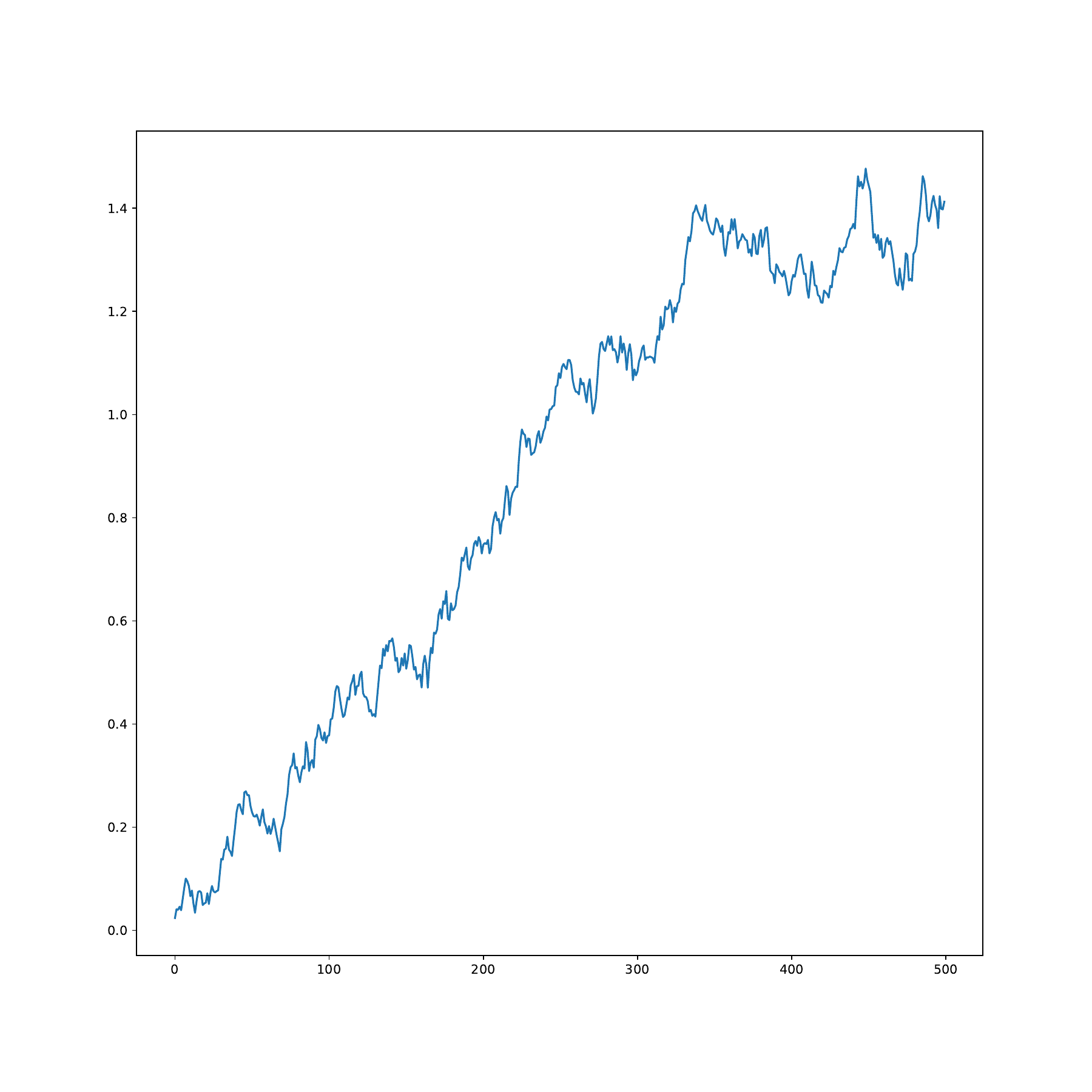}
  }
    \caption{Angle-axis representation of the guided bridge defined by \eqref{eq: guided diffusion sde}. (Left) The projection of the path in $\SO(3)$ to $\mathbb S^2$. The trajectory on $\mathbb S^2$ correspond to the motion of the tip of the blue vector as seen in Figure~\ref{fig: bridge on so3}. (Right) The angle representation of the guided bridge in $\SO(3)$.}
    \label{fig: angle-axis representation on SO(3)}
\end{figure}

In this section, we consider the task of simulating \eqref{eq: BM SDE on Lie group} conditioned to hit $v \in G$, at time $T > 0$. The potentially intractable transition density for the solution of \eqref{eq: BM SDE on Lie group} inhibits simulation directly from \eqref{eq: BB sde on frame bundle}. Instead, we propose to add a guiding term mimicking that of Delyon and Hu \cite{delyon_simulation_2006}, i.e., the guiding term becomes the gradient of the distance to $v$ divided by the time to arrival. The SDE for the guided diffusion becomes
    \begin{equation}\label{eq: guided diffusion sde}
        dY_t = - \frac{1}{2}V_0(Y_t) dt + V_i(Y_t) \circ \left(dB^i_t - \frac{\left(\nabla_{y|_{y=Y_t}} d(y,v)^2\right)^i}{2(T-t)}dt\right), \qquad Y_0 = e,
    \end{equation}
where $d(\cdot,v)$ denotes the Riemannian distance to $v$. Note that we can always, for convenience, take the initial value to be the identity $e$. Equation \eqref{eq: guided diffusion sde} can equivalently be written as
    \begin{equation*}
        dY_t = - \frac{1}{2}V_0(Y_t) dt + V_i(Y_t) \circ \left(dB^i_t - \frac{\Log_{Y_t}(v)^i}{T-t}dt\right), \qquad Y_0 = e,
    \end{equation*}
where $\Log_p$ is the inverse of the Riemannian exponential map $\Exp_p$. Figure~\ref{fig: bridge on so3} illustrates one sample path of the simulation scheme in \eqref{eq: guided diffusion sde} on the Lie group $\SO(3)$ and the corresponding axis-angle representation is visualized in Figure~\ref{fig: angle-axis representation on SO(3)}. 


The guiding term in \eqref{eq: guided diffusion sde} is identical to the guiding term described in \cite{jensen_simulation_2021}. However, in that case, the guided processes used the frame bundle of $M$. In the Lie group setting, since Lie groups are parallelizable, the use of the frame bundle is not needed. 

Numerical computations of the Lie group exponential map are often computationally efficient due to existence of certain algorithms (see \cite{pennec2019riemannian} and references therein). Therefore, by a change of measures argument, the equation above can be expressed in terms of the inverse of the Lie group exponential (this process is denoted $Y$ as well)
    \begin{equation}\label{eq: guided sde lie group log}
        dY_t = - \frac{1}{2}V_0(Y_t) dt + V_i(Y_t) \circ \left(d\bar B^i_t -\frac{\log_{Y_t}(v)^i}{T-t}dt\right)
    \end{equation}
$Y_0 = e$, where $\bar B$ is a Brownian motion under a new measure, say $\bar \Pb$. The measure $\bar \Pb$ can explicitly be expressed as
    \begin{equation}\label{eq: change to lie group log radon nikodym}
        \frac{d\bar \Pb}{d\Pb}|_{\cF_t}
        = 
        \exp\bigg[
        -
        \int_0^t H_v(s,Y_s) \nonumber
        \\
        - \frac{1}{2}
        \int_0^t \frac{ \lVert \log_{Y_s}(v)-\Log_{Y_s}(v)\rVert^2_{Y_s}}{(T-s)^2}ds \bigg],
    \end{equation}
where $\Pb$ denotes the law of the SDE in \eqref{eq: guided diffusion sde} and
\begin{equation*}
    H_v(t,Y_t)= \bigg\langle \frac{\left(\log_{Y_t}(v) 
        - \Log_{Y_t}(v)\right)}{T-t}, 
        V(Y_t)dB_t\bigg\rangle_{Y_t}.
\end{equation*}
The Radon-Nikodym derivative above is a martingale, whenever the group logarithm and the Riemannian logarithm coincide. This is for example the case when the metric is bi-invariant. 

\subsection{Radial process}
We now aim to investigate the relation between the bridge measure and the above simulation schemes.
Let $r_v(\cdot) := d(\cdot ,v)$ be the distance to $v$ such that $r_v(g_t)$ is the the radial process. Due to the singularities of the radial process on $\Cut(v) \cup \{v\}$, the usual It\^o's formula only applies on subsets away from the cut-locus. The extension beyond the cut-locus of a Brownian motion's radial process was due to Kendall \cite{kendall1987radial}. Barden and Le \cite{barden1997some, le1995ito} generalized the result to $M$-valued semimartingales. The radial process of the Brownian motion \eqref{eq: BM SDE on Lie group} is given by
    \begin{equation}\label{eq: radial process for BM}
        r_v(g_t) = r_v(g_0)^2 + \int_0^t \left\langle \nabla_{g_s} r_v(g_s), V(g_s)dB_s\right\rangle_{g_s} + \frac{1}{2}\int_0^t \Delta_G r_v(g_s) ds - L^v_s(g),
    \end{equation}
where $L^v$ is the geometric local time of the cut-locus $\Cut(v)$, which is non-decreasing continuous random functional increasing only when $g$ is in $\Cut(v)$ (see \cite{barden1997some,kendall1987radial,le1995ito}). Let $W_t := \int_0^t \left\langle \partiel r,  V_i(g_s) \right\rangle dB^i_s$, which is the local-martingale part in the above equation. The quadratic variation of $W_t$ satisfies
    $
        d[W,W]_t 
        = 
     dt,
   $
by the orthonormality of $\{V_1,\dots, V_d\}$, thus $W_t$ is a Brownian motion by Levy's characterization theorem. From the stochastic integration by parts formula and \eqref{eq: radial process for BM}, the squared radial process of $g$ satisfies 
    \begin{equation}\label{eq: squared radial process of BM}
         r_v(g_t)^2 = r_v(g_0)^2 + 2\int_0^t r_v(g_s) dW_s + \int_0^t r_v(g_s) \Delta_G r_v(g_s) ds - 2 \int_0^t r(g_s)dL^v_s,
    \end{equation}
where $dL^v_s$ is the random measure associated to $L^v_s(X)$.

Similarly, we obtain an expression for the squared radial process of $Y$. The radial process becomes
    \begin{equation}\label{eq: squared radial process of guided diffusion}
         r_v^2(g_t) = r_v(g_0)^2 + 2\int_0^t r_v(g_s) dW_s  + \int_0^t \frac{1}{2}\Delta_G r_v(g_s)^2 ds - \int_0^t \frac{r_v(g_s)^2}{T-s} ds - 2\int_0^t r_v(g_s) dL^v_s.
    \end{equation}
Imposing a growth condition on the radial process yields an $L^2$-bound on the radial process of the guided diffusion, \cite{thompson2018brownian}. So assume there exist constants $\nu \geq 1$ and $\lambda \in \R$ such that $\tfrac{1}{2} \Delta_G r_v^2 \leq \nu + \lambda r_v^2$ on $D\backslash \Cut(v)$, for every regular domain $D \subseteq G$. Then \eqref{eq: squared radial process of guided diffusion} satisfies
\begin{equation}\label{eq: mean squared radial bound} 
	\E [1_{t < \tau_D}r_v(Y_t)^2] \leq \left( r^2_v(e) + \nu t\left(\frac{t}{T-t}\right) \right)\left(\frac{T-t}{t}\right)^2 e^{\lambda t},
\end{equation}
where $\tau_D$ is the first exit time of $Y$ from the domain $D$.

\subsection{Girsanov change of measure}

Let $B_t$ be a $d$-dimensional Brownian motion defined on a filtered probability space $(\Omega, \cF, (\cF_s)_{s \geq 0},\Pb)$ and let $g_t$ be a solution of \eqref{eq: BM SDE on Lie group}. The process $\tfrac{\nabla r_v(g_t)^2}{2(T-t)}$ is an adapted process. As $g_t$ is non-explosive, we see that
    \begin{equation}\label{eq: novikov condition type bound on radial process}
    \int_0^t \left\lVert \frac{\nabla r_v(g_s)^2}{2(T-s)} \right\rVert^2 ds = \int_0^t \frac{r_v(g_s)^2 }{(T-s)^2}ds \leq C,
    \end{equation}
for every $0\leq t<T$, almost surely, and for some fixed constant $C > 0$. Define a new measure $\Qb$ by
    \begin{equation}\label{eq: radon-nikodym derivative}
       Z_t :=  \frac{d\Qb}{d\Pb}\bigg|_{\cF_t}(g)  = \exp\left[-\int_0^t \left\langle  \frac{\nabla r_v(g_s)^2}{2(T-s)}, V(g_t)dB_s\right\rangle - \frac{1}{2} \int_0^t \frac{r_v(g_s)^2 }{(T-s)^2}ds \right].
    \end{equation}
From \eqref{eq: novikov condition type bound on radial process}, the process $Z_t$ is a martingale, for $t \in [0, T)$, and $\Qb_t$ defines a probability measure on each $\cF_t$ absolutely continuous with respect to $\Pb$. By Girsanov's theorem (see e.g. \cite[Theorem 8.1.2]{hsu2002stochastic}), we get a new process $b_s$ which is a Brownian motion under the probability measure $\Qb$. Moreover, under the probability $\Qb$, equation \eqref{eq: BM SDE on Lie group} becomes 
    \begin{equation}\label{eq: guided diffusion sde with radial unit vector}
        dY_t = - \frac{1}{2}V_0(Y_t)dt + V_i(Y_t) \circ \left(db^i_t - \frac{r_v(Y_t)}{T-t} \left(\partiel{r_v}\right)^i dt\right),
    \end{equation}
where $\left(\partiel r\right)^i$ is the $i$'th component of the unit radial vector field in the direction of $v$. The squared radial vector field is smooth away from $\Cut(v)$ and thus we set it to zero on $\Cut(v)$. Away from $\Cut(v)$, the squared radial vector field is $2\Log_v$. The added drift term acts as a guiding term, which pulls the process towards $v$ at time $T>0$. 

From \eqref{eq: radon-nikodym derivative}, we see that $\E[f(Y_t)] = \E[f(g_t)Z_t]$. Using \eqref{eq: squared radial process of BM} and the identity $\Delta_G r_v = \frac{d-1}{r_v} +  \partiel {r_v} \log \Theta_v$, $\theta_v$ being the Jacobian determinant of $\Exp_v$ (see e.g. \cite{thompson2015submanifold}), we equivalently write $\E[f(Y_t)\varphi_t] = \E[f(X_t)\psi_t]$, with
    \begin{equation}\label{eq: psi and phi function}
        \psi_{t,v} 
        := 
        \exp\left[\frac{-r_v^2(g_t)}{2(T-t)}\right]
        \qquad
        \varphi_{t,v}
        := 
        \exp\left[ \int_0^t \frac{r_v(Y_s)}{T-s}\left(dA^v_s + dL^v_s\right)\right],
    \end{equation}
where 
$dA^v_s = \partiel {r_v} \log \theta_v^{-1/2}(Y_s)ds$ is a random measure supported on $G \backslash \Cut(v)$, and $dL_s^v$ the geometric local time at $\Cut(v)$.

\subsection{Delyon and Hu in Lie groups}

We can now generalize the result of Delyon and Hu \cite[Theorem 5]{delyon_simulation_2006} to the Lie group setting. 
The result here for Lie groups is analogous to the Riemmanian setting as covered in \cite{jensen_simulation_2021}.

    \begin{Theorem}\label{result: delyon and hu result}
    Let $g_t$ be a solution of \eqref{eq: BM SDE on Lie group}. The SDE \eqref{eq: guided diffusion sde} yields a strong solution on $[0,T)$ and satisfies $\lim_{t\uparrow T} Y_t = v$ almost surely. Moreover, the conditional expectation of $g$ given $g_T=v$ is
    \begin{equation}\label{eq: delyon hu on lie group}
        \E[f(g)|g_T = v] = \lim_{t\uparrow T} \frac{\E\left[f(Y) \varphi_{t,v}\right]}{\E\left[ \varphi_{t,v}\right]},
    \end{equation}
    for every $\cF_t$-measurable non-negative function $f$ on $G$, for $t\in [0,T)$ where $\varphi_t$ is given in \eqref{eq: psi and phi function}.
    \end{Theorem}
    
When the geometry of $G$ is particularly simple the equivalence of measures hold on $[0,T]$, see \cite{jensen_simulation_2021}, and the above result reduces to the following.
    
    \begin{Corollary}
    When $G$ is simply connected, \eqref{eq: delyon hu on lie group} becomes
    \begin{equation}
        \E[f(g)|g_T = v] = C \E\left[f(Y) \varphi_{T,v}\right],
    \end{equation}
    where $C > 0$ is a constant, which depends on the initial point, the time $T > 0$, and the curvature in the radial direction.
    \end{Corollary}

\section{Simulation of bridges in homogeneous spaces}\label{sec:simulation of bridges on Lie groups and Homogeneous spaces}

We now turn to bridge simulation in homogeneous spaces by sampling bridges in $G$ conditioned on the fiber over $v \in M = G/H$. 
We simulate in the top space, that is, bridge simulation schemes on the Lie group $G$, and subsequently project to the homogeneous space $M$. We will be considering two schemes. Let $v \in M$.

\begin{enumerate}
\item Find closest point $\bar v$ in the fiber $\pi^{-1}(v)v$ and iteratively update $\bar v$ at each time step. 
%
%
  \item Sample $k$-points $\{\bar v_1, \dots, \bar v_k\}$ in fiber above $v \in M$ and consider the bridge $Y$ in $G$ conditioned on $Y_T \in \{\bar v_1,\dots , \bar v_k\}$. 
\end{enumerate}

The motivation for considering the two schemes above derives from  \cite{jensen2019simulation}. Here, the simulation of Brownian bridges on the flat torus $\mathbb T^2 = \R^2/ \mathbb Z^2$ was the focal point. In this particular geometrical context of $\mathbb T^2$, the lift of a $\mathbb T^2$-valued Brownian bridge results in a $\R^2$-valued Brownian bridge conditioned on a set $N \cong \mathbb Z^2$. The first scheme was considered in \cite{jensen2019simulation}, where Brownian bridges on the flat torus $\mathbb T^2 = \R^2/ \mathbb Z^2$ were lifted to bridges in $\R^2$. The second scheme provided a truncated version of the true bridge, for a suitable choice of $k$. Such a choice of $k$ is dependent on the time to arrival $T>0$ and the diffusivity $\sigma$. 

We can sample the $k$ points in the fiber $N=\pi^{-1}(v)$. However, we need to specify from which distribution in $N$ we sample. One approach to defining a distribution on $N$ uses the transition density of a Brownian motion. Thus, if there exists a point $v \in N$ closest to $g_0 \in G$, i.e., $d(v,g_0) \leq d(y,g_0)$, for all $y \in N$, then recording the endpoint $B_T$ of a Brownian motion $B_t$ in $N$, started at $B_0 = v$ corresponds to sampling from a normal distribution in $N$. As the time $T$ increases, the distribution tends to be uniform, and the initial starting point becomes irrelevant. Therefore, if no unique point $v \in N$ closest to $g_0$ exists, sampling from a uniform distribution seems more appropriate.

\subsection{Guiding to closest point}\label{sec: guide to nearest point}

	
%
%


Recall that the projection $\pi \colon G \rightarrow G/H$ is a submersion, hence the manifold $M = G/H$ is an embedded submanifold of $G$. From Lemma~\ref{lemma: conditional expectation on M}, we obtain a conditional expectation in $M$ by conditioning on the fiber in the Lie group. The corresponding SDE for the Fermi bridge in the Lie group setting is given by
\begin{equation}\label{eq: guided sde to fiber}
    dY_t = -\frac{1}{2}V_0(Y_t)dt + V_i(Y_t) \circ \left(dB_t^i -  \frac{\left(\nabla_{y|y=Y_t}d(y,N)^2\right)^i}{2(T-t)}dt \right), \quad Y_0 = e,
\end{equation}
where $d(x,N) := \inf_{z\in N} d(x,z)$ and $N := \pi^{-1}(v)$, for some $v \in M$.

The one-point motion conditioned on $v\in M$ corresponds to conditioning $g_t$ on the fiber $N := \pi^{-1}(v)$, and we can use Fermi bridges directly. Because $N$ is an embedded submanifold of $G$, we get from Thompson \cite{thompson2018brownian} that $\varphi_{t,N}$ is of the form
\begin{equation}\label{eq: phi term submanifold}
    \varphi_{t,N}
        := 
        \exp\left[ \int_0^t \frac{r_N(Y_s)}{T-s}\left(dA^N_s + dL^N_s\right)\right],
\end{equation}
where $dA^N_s =  \partiel {r_N} \log \Theta_N^{-1/2}(Y_s)ds$ and $\Theta_N = \theta_N \circ \left(\Exp|_{\Log\left(M\backslash \Cut(N)\right)}\right)^{-1}$. 
Similar to the single point case, we obtain
\begin{equation*}
    \E[f(X)|X_T \in N] = \lim_{t \uparrow T} \frac{\E[f(Y)\varphi_{t,N}]}{\E[\varphi_{t,N}]},
\end{equation*}
for any bounded measurable function $f$. There are various occasions where it can be justified to take the limit inside. See the discussion in \cite[Appendix C]{thompson_submanifold_2015}.

\subsection{Guiding to k-points in fiber}\label{sec: guide to k-points}

The guiding scheme presented in \eqref{eq: guided sde to fiber} requires an optimization step in each time step, and for practical purposes, this might be computationally inefficient. This section suggests guiding to a subset of the fiber $N$, reducing the guided bridge scheme in \eqref{eq: guided sde to fiber} to a case of guiding to a finite set of points.

For certain homogeneous spaces, e.g., $\mathbb T^2 = \R^2/\mathbb Z^2$, the fiber $N = \pi^{-1}(v)$ is a discrete subgroup in $G$. In this case, the volume measure $\Vol_N$ in \eqref{eq: density on submanifold} is the counting measure and we can write the density as
    \begin{equation*}
        p_t^G(g,N) = \sum_{v \in N} p^G_t(g,v).
    \end{equation*}
From a numerical perspective, when the discrete subgroup is large restricting to a smaller finite subgroup $N_k \subseteq N$ of $k$-nearest-points of the initial starting point may speed up computation-time. 

\begin{Example}
To exemplify this, let $p_t$ denote the transition density of a standard Brownian motion in $\R^2$. The transition density of the Brownian motion started at the origin is then given as $p_t(0,v) = \left(2 \pi t\right)^{-1}\exp\left(-\tfrac{\lVert v \rVert^2}{2t}\right)$. Let $N_k = \{(-2,-2),\dots, (2,2)\} \subseteq \mathbb Z^2$ be the subset of five-by-five grid points and let $T=1$, then we see 
    \begin{equation*}
         p_T^G(0,N) \approx \sum_{v \in N_k} p_T(0,v) = \left(2\pi\right)^{-1}\left(1 + 4\left(e^{-1/2} + e^{-1} + e^{-2} + e^{-4} +2e^{-8}\right) \right) \approx 0.98.
    \end{equation*}
We recover more than $98\%$ of the total mass when restricting to a finite set of points. If we restricted the subset to the set of three-by-three grid points $N_k = \{(-1,-1),\dots, (1,1)\}$, the density will only describe roughly $78\%$ of the total mass. However, if we restricted the time to arrival $T=1/2$, then we would recover $95\%$ of the mass. Thus, we see that both the initial point and the terminal time will affect the choice of $k$.
\end{Example}

The tractable transition density on the flat torus $\mathbb T^2 = \R^2 /\mathbb Z^2$ is given by $p^{\R}_T(0,N) = \sum_{v_i \in N} p^{\R}_T(0,v_i)$, where $N$ is a set isomorphic to $\mathbb Z^2$. The corresponding SDE for the conditioned process is given by \cite{jensen2019simulation}
\begin{equation}
    dY_t = \sum_{v_i \in N} f_i(t,Y_t) \frac{v_i-Y_t}{T-t}dt + dW_t, \quad \text{where} \quad f_i(t,Y_t) = \frac{\exp\left(- \frac{\lVert v_i - Y_t\rVert^2}{2t}\right)}{\sum_{v_j\in N}\exp\left(-\frac{\lVert v_j -Y_t\rVert^2}{2t}\right)}.
\end{equation}
By the example above, good numerical approximations can be obtained by restricting to a finite set of points. We conjecture that a similar type of guided drift can be used, where the transition density above is exchanged with the transition density of the Riemannian normal distribution (see e.g. \cite{thompson_submanifold_2015}). We do not pursue this approach any further in this paper. Instead, we propose to sample a point from $N$ from a given distribution on $N$.

    The following result is inspired by the type of conditioning found in van der Meulen and Schauer \cite{van2018bayesian}, Mider, Schauer, and van der Meulen \cite{mider2021continuous}, and Arnaudon et al. \cite{arnaudon2022diffusion}. We adopt this type of inexact matching by imposing noise on the conditioning point. The proposed method alleviates the optimization procedure in each time step to finding the closest point in the fiber. One immediate application of the result below will be in the situation where we sample Brownian motions in the fiber, starting at the closest point in the fiber. Recording the endpoints after some fixed time, we obtain samples from a normal distribution in the fiber. Therefore, the simulation scheme reduces to conditioning at a point as described in Section~\ref{sec: simulation of bridges on Lie groups}, the caveat being that the endpoint is tilted to a specific distribution. The result below is a theoretical one. We note from the guided bridge scheme that $d\Pb_{x,v}^T/d\Pb_x(y) = \varphi_{T,v}/\E[\varphi_{T,v}]$. 
    
    \begin{Theorem}
        Let $g$ be a Markov process defined on $(\Omega, \cF, \Pb)$ with values in $G$, and let $p_{t}^G(\cdot, \cdot)$ be its transition density defined by $\Pb(g_T \in du |g_t = g) = p^G_{T-t}(g,u)d\Vol_G(u)$. Assume $0 \leq t < T$ and $g_T \sim f_{u_0} \cdot \Vol_N$ (e.g., normal/uniform distribution on fiber over $v$) under the probability measure $\Pb_{g_0}$ started at $g_0$. 
        The conditional law of $g_t$ given $g_T=v$, $\Pb^T_{g_0,v}$, has density wrt. the reference measure $d\Vol_G$ given by
            \begin{equation}\label{eq: conditional density of brownian bridge}
                \frac{p^G_{T-t}(z,v)p^G_t(z,g_0)}{p^G_T(g_0,v)},
            \end{equation}
        and the simultaneous distribution of $(X_t,X_T)$ has density given by      \begin{equation}\label{eq: simulataneous density for x_t and x_T}
                \frac{\Pb(g_t \in dg,g_T \in du )}{d\Vol_N(u) d\Vol_G(g)} = f_{u_0}(u) \frac{p^G_{T-t}(z,u)p^G_t(z,g)}{p^G_T(g_0,u)}.
            \end{equation}
         Furthermore, if we define the $h$-function as
            \begin{equation}\label{eq: h function for X_T = f}
                h(t,g_t) = \frac{\int f_{u_0}(u) \frac{p^G_{T-t}(g_t,u)}{p_T^G(g_0,u)}d\Vol_N(u)}{\int_N f_{u_0}(x)d\Vol_N(x)},
            \end{equation}
        then for any non-negative measurable functional $F$ we have
            \begin{equation*}
                \E_{\Qb}[F(g)] 
                =
                \E_\Pb[h(t,g_t) F(g)] 
                =
                \int \E_\Pb[F(g)|g_T = u] \frac{f_{u_0}(u)}{\int_N f_{u_0}(x) d\Vol_N(x)} d\Vol_N(u).
            \end{equation*}
         The conditional distribution of $X_t$ given $X_T$ has density
            \begin{equation}
                k_t(g,u) = \frac{f_{u_0}(u)}{\int_N f_{u_0}(x)d\Vol_N(x)} \frac{p^G_{T-t}(z,u)p^G_t(z,g)}{p^G_T(g_0,u)},
            \end{equation}
        with respect to the volume measure $\Vol_N$.
         If the distribution $g_T(\Pb)$ has full mass in the fiber $N$, i.e., $g_T(\Pb)(N) = \int_N f dVol =1$ the term above simplifies.
    \end{Theorem}
    
    \begin{proof}
    The fact that \eqref{eq: conditional density of brownian bridge} is the conditional density wrt. $d\Vol_G$, the volume measure on $G$, follows from \eqref{eq: radon-nikodym for v}, since $d\Pb^t_g = p_t(g, z)d\Vol_G(z)$ and therefore
        \begin{equation*}
            d\Pb^T_{g_0,v}(g_t) = \frac{p^G_{T-t}(z,v)p^G_t(g_0,z)}{p^G_T(g_0,v)}d\Vol_G(z).
        \end{equation*}
      Hence \eqref{eq: simulataneous density for x_t and x_T} follows.
      
       For the second part take $h$ as defined in \eqref{eq: h function for X_T = f}. Without loss of generality assume that $g_T(\Pb)(N)=1$. Note that $h$ is a martingale with $h(0,g_0) =1$, since $g$ is a Markov process and
            \begin{align*}
                \E[h(t,g_t)|g_s] 
                =
                &
                \int p^G_{t-s}(g_s,z) h(t,z) d\Vol_G(z)
                \\
                =
                &
                \int p^G_{t-s}(g_s,z) \int f_{u_0}(x) \frac{p^G_{T-t}(z,x)}{p_T^G(g_0,x)}d\Vol_N(x) d\Vol_G(z)
                \\
                =
                &
                \int f_{u_0}(x) \frac{p^G_{T-s}(g_s,x)}{p_T^G(g_0,x)}d\Vol_N(x) = h(s,g_s)
            \end{align*}
        together with 
            \begin{equation*}
                \E[h(t,g_t)] =  \int f_{u_0}(x) \frac{p^G_{T}(g_0,x)}{p_T^G(g_0,x)}d\Vol_N(x) = 1.
            \end{equation*}
        Since $\lim_{t \downarrow 0} \int p^G_t(g_0,z)f(z)d\Vol_G(z) =f(g_0)$, for any bounded continuous function $f$, Fatou's lemma ensures that $\E[h(T,g_T)]= 1$ 
            \begin{align*}
                1 
                & 
                =
                \limsup_{t\uparrow T} \E[h(t,g_t)] 
                \leq
                \E\left[\limsup_{t\uparrow T} h(t,g_t)\right]
                \\
                &
                =
                \int_G \limsup_{t \uparrow T} \int f_{u_0}(x) \frac{p^G_{T-t}(z,x)}{p_T^G(g_0,x)}d\Vol_N(x) d\Vol_G(z)
                \\
                &
                =
                \int_G \frac{f_{u_0}(z)}{p_T^G(g_0,z)} d\Vol_G(z)
                =
                \int_G \liminf_{t \uparrow T} \int f_{u_0}(x) \frac{p^G_{T-t}(z,x)}{p_T^G(g_0,x)}d\Vol_N(x) d\Vol_G(z)
                \\
                &
                \leq
                \liminf_{t \uparrow T} \E[h(t,g_t)] = 1.
            \end{align*}
        Hence $h$ is a true martingale on $[0,T]$ and thus defines a new probability measure $\Qb$ on $\cF$ by $\frac{d\Qb}{d\Pb}|_{\cF_t}(g) = h(t,g_t)$. 
        
        For the second part of the proof, assume, temporarily, that $F$ is a measurable function such that
            \begin{align*}
                \E_\Pb[h(t,g_t) F(g_t)] 
                &
                =
                \int  p^G_{t}(g_0,z)  h(t,z)F(z) d\Vol_G(z)
                \\
                &
                = 
                \int \int  \frac{p^G_{T-t}(z,x)p^G_t(g_0,z)}{p_T^G(g_0,x)} F(z)d\Vol_G(z) f(x) d\Vol_N(x)
                \\
                &
                =
                \int \int F(z) d\Pb_{g_0,x}^T(z) f_{u_0}(x) d\Vol_N(x)
                =
                \int \E_\Pb[F(g_t)|g_T = x] f_{u_0}(x) d\Vol_N(x).
            \end{align*}
        In order to conclude, we need to show that for any finite distribution $(X_{t_1},\dots, X_{t_n})$ 
            \begin{equation*}
                \E_\Pb[h(t,X_t) F(X_{t_1},\dots, X_{t_n})] =  \int \E_\Pb[F(X_{t_1},\dots, X_{t_n})|X_T = u] f_{u_0}(u) d\Vol_N(u).
            \end{equation*}
        Therefore, let $0<t_1<\dots < t_n < T$ and $t \in (t_n,T)$. Define $\Phi_F$ similar to how it was defined in \cite{jensen_simulation_2021}. Then 
            \begin{align*}
                \E_\Pb[h(t,X_t) F(X_{t_1},\dots, X_{t_n})]
                =
                &
                \int_G h(t,x) \Phi_F(t,x) d\Vol_G(x)
                \\
                =
                &
                \int f_{u_0}(u)\int F(z)P(z,x,u) d\Vol(z) d\Vol(x)d\Vol(u)
                \\
                =
                & 
                \int \E_\Pb[F(X_{t_1},\dots, X_{t_n})|X_T = u] f_{u_0}(u) d\Vol_N(u),
            \end{align*}
        where $z=(z_1,\dots, z_n)$ and $d\Vol(z) = d\Vol(z_1)\dots d\Vol(z_n)$ and where
            \begin{equation*}
                P(z,x,u) = \frac{p^G_{t_1}(x_0,z_1)\dots p^G_{t-t_n}(z_n,x)p^G_{T-t}(x,u)}{p_T^G(x_0,u)}.
            \end{equation*}
    \end{proof}

Assume that the constants $c_i$ exist. We can obtain estimates of the constants $c_i$ via simulation of sample paths of the guided process. Let $y^i_t = Y_t(\omega_i)$ be realizations of the guided bridge process. We can then use the estimator 
    \begin{equation}
        \bar \varphi_{T,v_i}(y) =\frac{1}{m} \sum_{n=1}^m  \exp\left[\int_0^T \frac{r_{v_i}(y^n_s)}{T-s}\left(dA^{v_i}_s + dL_s^{v_i}\right)\right]
    \end{equation}
to approximate the constants $c_i = \E[\varphi_{T,v_i}]$. Algorithm~\ref{alg: stoch metropolis-hastings} provides a method for obtaining $k$-points in the fiber $N$, together with the normalizing constants, when $N$ is connected and compact.

\begin{algorithm} \DontPrintSemicolon \SetAlgoLined
\tcp{Initialization} Choose initial point $e \in G$ and $v_1\in N$ closest to $e$. Simulate a guided bridge process to $v_1$ and obtain an estimate of $\E[\varphi_{T,v_1}]$.
\\
\tcp{Main loop}
\While{$k$ points not reached}{ 
  \tcp{Step 1:}
  Propose $u$ from the proposal density $f_{v_i}(u)$ (e.g. uniform density in $N$ centered at $v_i$ or normal density in $N$ centered at $v_i$) and sample estimator for $\E[\varphi_{T,u}]$
  \\
  \tcp{Step 2:}
  Calculate the acceptance ratio 
$g(u,v_i) 
= 
\min
\left\{1,
\frac{ f_u(v_i) q_{T}(x_0,u)
\bar \varphi_{T,u}
}
{ 
f_{v_i}(u) q_{T}(x_0,v_i)\bar \varphi_{T,v_i}
}
\right\}
$
(Note that if $f$ is symmetric it cancels out in the acceptance probability)
\\
\tcp{Step 3} 
Accept with probability g$(u,v_i)$ and set $v_{i+1} = u$ as well as $c_{i+1} = \bar \varphi_{T,v_{i+1}}$ otherwise do nothing.
}
\tcp{Output:} 
$\{(v_1,c_1), \dots , (v_k,c_k)\}$
\caption{Stochastic Metropolis-Hastings Algorithm}
\label{alg: stoch metropolis-hastings}
\end{algorithm}

\section{Maximum Likelihood Estimation}\label{sec: MLE}

For a manifold-valued Brownian motion recall \eqref{eq: brownian motion local}, which describes the Brownian motion locally in a chart. The diffusion coefficient is the matrix square root of the cometric tensor, i.e., the inverse metric tensor. As seen in Figure~\ref{fig: anisotropic distribution on S2}, the pushforward measure of a Brownian motion generated by a non-invariant metric induces anistotropic distributions on the quotient space. We aim to estimate the underlying metric by an MLE approach.

Consider observations $y^1,\dots, y^n$ on $G$ or $G/H$ obtained from distributions $\Pb_\theta^t$ or $\pi_*\Pb_\theta^t$, with parameters $\theta = (g,A)$, and corresponding densities $p_t(\cdot|\theta)$ and $\pi_*p_t(\cdot|\theta)$. Here $A^{-1} = \Sigma$ is the inverse covariance matrix and $\Sigma = \sigma \sigma^T$. We can define the likelihood function as 
\begin{equation}\label{eq: likelihood MLE}
    \mathcal L(\theta| y^1,\dots, y^n;T) = \prod_{i=1}^n p_T( y^i|\theta),
\end{equation}
and similarly $\pi_* \mathcal L = \prod \pi_* p$. The bridge sampling scheme introduced above yields approximations of the intractable transition densities in \ref{eq: likelihood MLE}. Algorithm~\ref{alg: iterative maximum likelihood} provides a detailed description for the iterative MLE approach. Visual examples of the iterative MLE can be found in Figure~\ref{fig: Metric estimation on SO(3)} and \ref{fig: diffusion mean estimation on gl3 and spd3}.

\begin{algorithm} \DontPrintSemicolon \SetAlgoLined
\caption{Parameter Estimation: Iterative MLE.}
\label{alg: iterative maximum likelihood}
\tcp{Initialization}
Given $n$ data points $\{v_1,\dots, v_n\}$.\\ \
\tcp{Specify initial parameters $\theta_0 = (g_0, A_0)$ and a learning rate $\eta$.}
\For{$k=1$ to $K$}
    {
        \For{$j=1$ to $n$}
        {
        Sample $m$ bridges from \eqref{eq: guided diffusion sde} conditioned on $v_j$ to get estimate for\ $\E[\varphi_{T,v_j}] \approx \frac{1}{m}\sum_{i=1}^m \varphi_{T,v_j}^i$
        }
        $\ell_{\theta_{k-1}}(v_1,\dots, v_n) = \prod_{j=1}^n \left(\frac{\det A_{k-1}(T,v_j)}{2\pi T}\right)^{3/2} e^{-\frac{\lVert \Log_{v_j}(g_{k-1})\rVert^2_{A_{k-1}} }{2T}} \frac{1}{m}\sum_{i=1}^m \varphi_{T,v_j}^i$\\ \
        \tcp{Compute the gradient}\
        $\upxi_{k} = \nabla_{\theta_{k-1}} \log \ell_{\theta_{k-1}}(v_1,\dots, v_n)$\\ \
        \tcp{Update the parameters}\;
        $\theta_k = \theta_{k-1} - \eta \upxi_k$\;
    }
\tcp{Return final parameters $\theta_K = (g_K, A_K)$}
\end{algorithm}

\section{Experiments}\label{sec: numerical experiments}

In this section, we present numerical results of bridge sampling on specific Lie groups and homogeneous spaces. The specific Lie groups in question are the three-dimensional rotation group $\SO(3)$ and the general linear group of invertible matrices with positive determinant $\GL_+(3)$. 
Exploiting the bridge sampling scheme described above, we show below how to estimate the true underlying metric on $\SO(3)$ with iterative maximum likelihood estimation. 

The space of the symmetric positive definite matrices $\SPD(n)$ is an example of a non-linear space in which geometric data appear in many applications. The space $\SPD(3)$ can be obtained as the homogeneous space $\GL_+(3)/\SO(3)$, where $\GL_+$ is the space of invertible matrices with a positive determinant. 

Lastly, considering the two-sphere $\mathbb S^2$ as the homogeneous space $\SO(3)/\SO(2)$, we verify that the bridge sampling scheme on this homogeneous space yields admissible heat kernel estimates on $\mathbb S^2$.

\subsection{Numerical simulations} 
The Euler-Heun scheme leads to approximation of the Stratonovich integral. With a time
discretization $t_1,\ldots,t_k$, $t_k-t_{k-1}=\Delta t$ and corresponding noise
$\Delta B_{t_i}\sim N(0,\Delta t)$, the numerical approximation of the Brownian motion \eqref{eq: BM SDE on Lie group} takes the form
\begin{equation}
  x_{t_{k+1}} = 
  x_{t_k}
  -
  \frac{1}{2}\sum_{j,i}C^j_{ij}V_i(x_{t_k})\Delta t
  +
  \frac{v_{t_{k+1}}+V_i(v_{t_{k+1}}+x_{t_{k}})\Delta B_{t_k}^i}{2}
\label{eq: numerical brownian motion}
\end{equation}
where $v_{t_{k+1}} = V_i(x_{t_k})\Delta B_{t_k}^i$ is only used as an intermediate value in integration. Adding the logarithmic term in \eqref{eq: guided diffusion sde with radial unit vector} to \eqref{eq: numerical brownian motion} we obtain a numerical approximation of a guided diffusion \eqref{eq: guided diffusion sde}.

\subsection{Importance sampling and metric estimation on $\SO(3)$}\label{sec: importance sampling and metric estimation}


This section takes $G$ as the special orthogonal group $\SO(3)$, the space of three-dimensional rotation matrices. The special orthogonal group is a compact connected matrix Lie group. In the context of matrix Lie groups, computing left-invariant vector fields is straightforward. The Lie algebra of the rotation group $\SO(3)$ is the space of three-by-three skew symmetric matrices, $\mathfrak{so}(3)$. The exponential map $\exp \colon \mathfrak{so}(3) \rightarrow \SO(3)$ coincides with the usual matrix exponential $e^A$. With $a \in \R^3$, we can express any element $A \in \mathfrak{so}(3)$ in terms of the standard basis $\{e_1, e_2, e_3\}$ of $\R^3$ as
\begin{equation*}
   A =  \begin{pmatrix}
    0 & -a_3 & a_2\\
    a_3 & 0 & -a_1 \\
    -a_2 & a_1 & 0
    \end{pmatrix}.
\end{equation*}
Let $\theta = \lVert a \rVert_2$ and assume that $\theta \neq 0$. By Rodrigues' formula the matrix Lie group exponential map $\exp \colon \mathfrak{so}(3) \rightarrow \SO(3)$ is given by
    \begin{equation*}
        R:= 
        e^A = I + \frac{\sin(\theta)}{\theta}A
        +
        \frac{(1-\cos(\theta))}{\theta^2}A^2
    \end{equation*}
and the corresponding inverse matrix Lie group exponential map $\log \colon \SO(3) \rightarrow \mathfrak{so}(3)$ 
    \begin{equation*}
        \log(R) = \frac{\sin^{-1}(\theta)}{2\theta}(R - R^T).
    \end{equation*}
The rotation group $\SO(3)$ is a semi-simple Lie group; hence, a bi-invariant inner product exists. In the case of a bi-invariant metric, the Riemannian exponential map $\Exp$ coincides with the Lie group exponential map $\exp$ and thus the Riemannian distance function $d(R,I)^2 = \lVert \Log_I(R) \rVert^2$, from the rotation $R$ to the identity $I$, satisfies $\nabla_R d(R,I)^2 = 2 \log(R)$.

The structure coefficients of $\mathfrak{so}(3)$ are particularly simple. Let $A_i = A$ with $a_j =1$ if $i=j$ and zero otherwise. In this case, $\{A_1, A_2, A_3\}$ defines a basis of $\mathfrak{so}(3)$. The structure coefficients satisfy the relation $[A_i, A_j] = C^k_{ij}A_k = \epsilon^{ijk}A_k$, where $\epsilon^{ijk}$ denotes the Levi-Civita symbols. The Levi-Civita symbols are defined as $+1$, for $(i,j,k)$ an even permutation of $(1,2,3)$, $-1$ for every odd permutation, and zero otherwise.

\subsection{Numerical bridge sampling algorithm on $\SO(3)$}

Utilizing the simple expressions for the structure coefficients and the Lie group logarithmic map, we can explicitly write up the numerical approximation of the guided bridge processes (Brownian bridge) on $\SO(3)$ as

\begin{equation}
  x_{t_{k+1}} = 
  x_{t_k}
  -
  \frac{1}{2}\sum_{j,i}\epsilon^{ijj}V_i(x_{t_k})\Delta t
  +
  \frac{v_{t_{k+1}}+V_i(v_{t_{k+1}}+x_{t_{k}})\left(\Delta B_{t_k}^i - \frac{\log(x_k)}{T-t_k}\Delta t \right)}{2},
\end{equation}
where in this case we have $v_{t_{k+1}} = V_i(x_{t_k})\left(\Delta B_{t_k}^i- \frac{\log(x_k)}{T-t_k}\Delta t\right)$.
Figure \ref{fig: bridge on so3} illustrates the numerical approximation by showcasing three different sample paths from the guided diffusion conditioned to hit the rotation represented by the black vectors.

Another way of visualizing the guided bridge on the rotation group $\SO(3)$ is through the angle-axis representation. Figure \ref{fig: angle-axis representation on SO(3)} represents a guided process on $\SO(3)$ by presenting the axis representation on $\mathbb S^2$ and its corresponding angle of rotation.

\subsection{Metric estimation on the three-dimensional rotation group}
 In the $d$-dimensional Euclidean case, importance sampling yields the estimate \cite{papaspiliopoulos2012importance}
    \begin{equation*}
        p_T(u,v) = \left(\frac{\det\left(A(T,v)\right)}{2 \pi T}\right)^{\tfrac{d}{2}} e^{-\frac{\lVert u-v\rVert^2_A}{2T}} \E[\varphi_{T,v}],
    \end{equation*}
where $\lVert x \rVert_A = x^TA(0,u)x$. Thus, from the output of the importance sampling, we get an estimate of the transition density.
Similar to the Euclidean case, we obtain an expression for the heat kernel $p_T(e,v)$ as $p_T(e,v) = q(T,e)  \E\left[\varphi_{T,v}\right]$, where
	\begin{equation}\label{eq: q function in density}
	q(T,g) 
	= 
	\left(
	\frac{
	\det A(v)
	}
	{
	2\pi T
	}
	\right)^{\tfrac{3}{2}} 
	\exp
	\left(
	-\frac{d(g,v)^2}{2T}
	\right) 
	= 
	\left(\frac{\det A(T,v)}{2\pi T}\right)^{\tfrac{3}{2}} \exp\left(-\frac{\lVert \Log_g(v)\rVert^2_A }{2T}\right),
	\end{equation}
where the equality holds almost everywhere, and $A \in \Symp(\cG)$ denotes the metric $A(e): = A(0,e)$. The $\Log_g$ map in \eqref{eq: q function in density} is the Riemannian inverse exponential map $(\Exp_g)^{-1}$. 

\begin{figure}[H]
  \subfloat{
        \includegraphics[width=.40\linewidth,clip=true,trim=50 50 50 50]{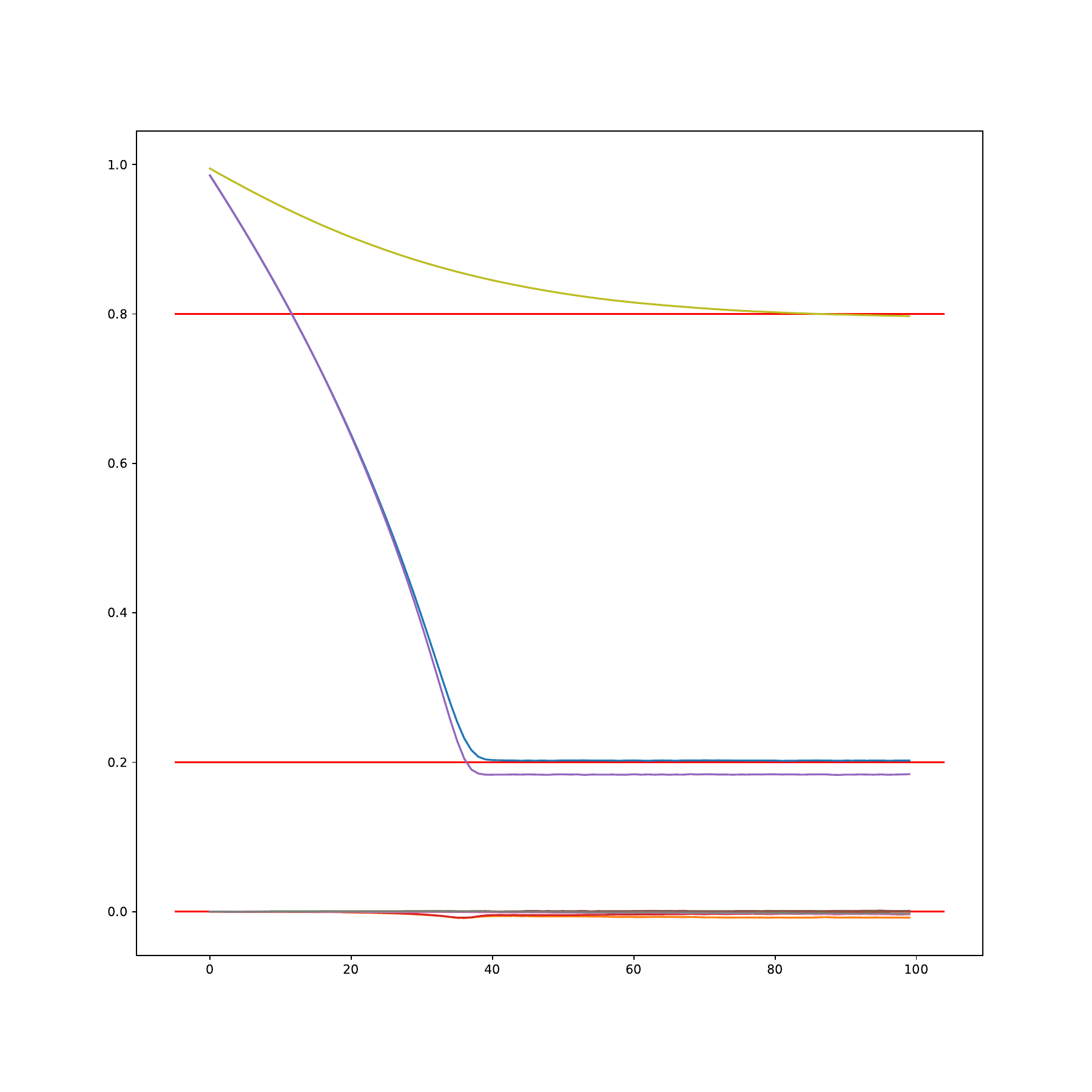}
      }
    \hfill
    \subfloat{
    \includegraphics[width=.40\linewidth,clip=true,trim=50 50 50 50]{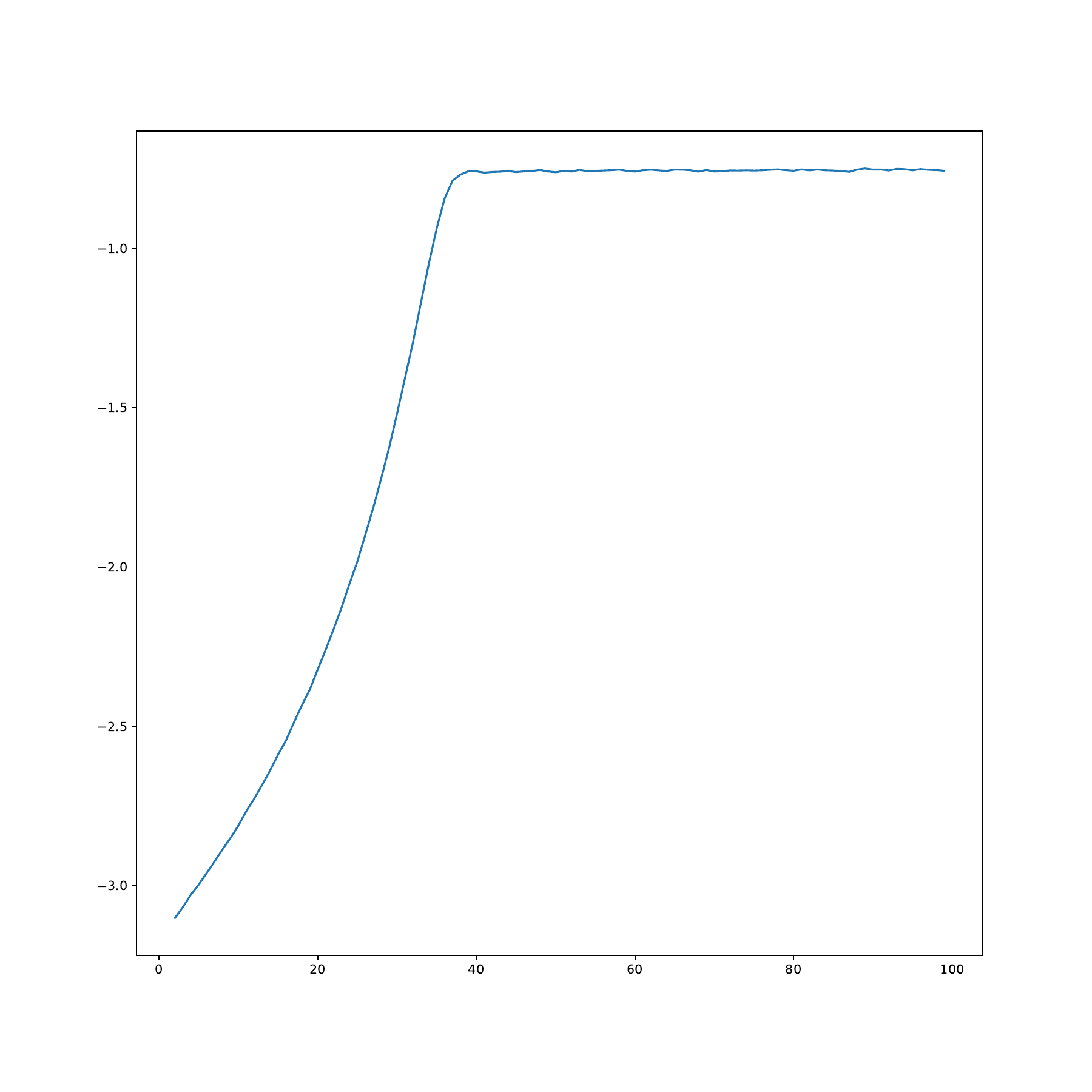}
  }
   \hfill
    \subfloat{
    \includegraphics[width=.40\linewidth,clip=true,trim=50 50 50 50]{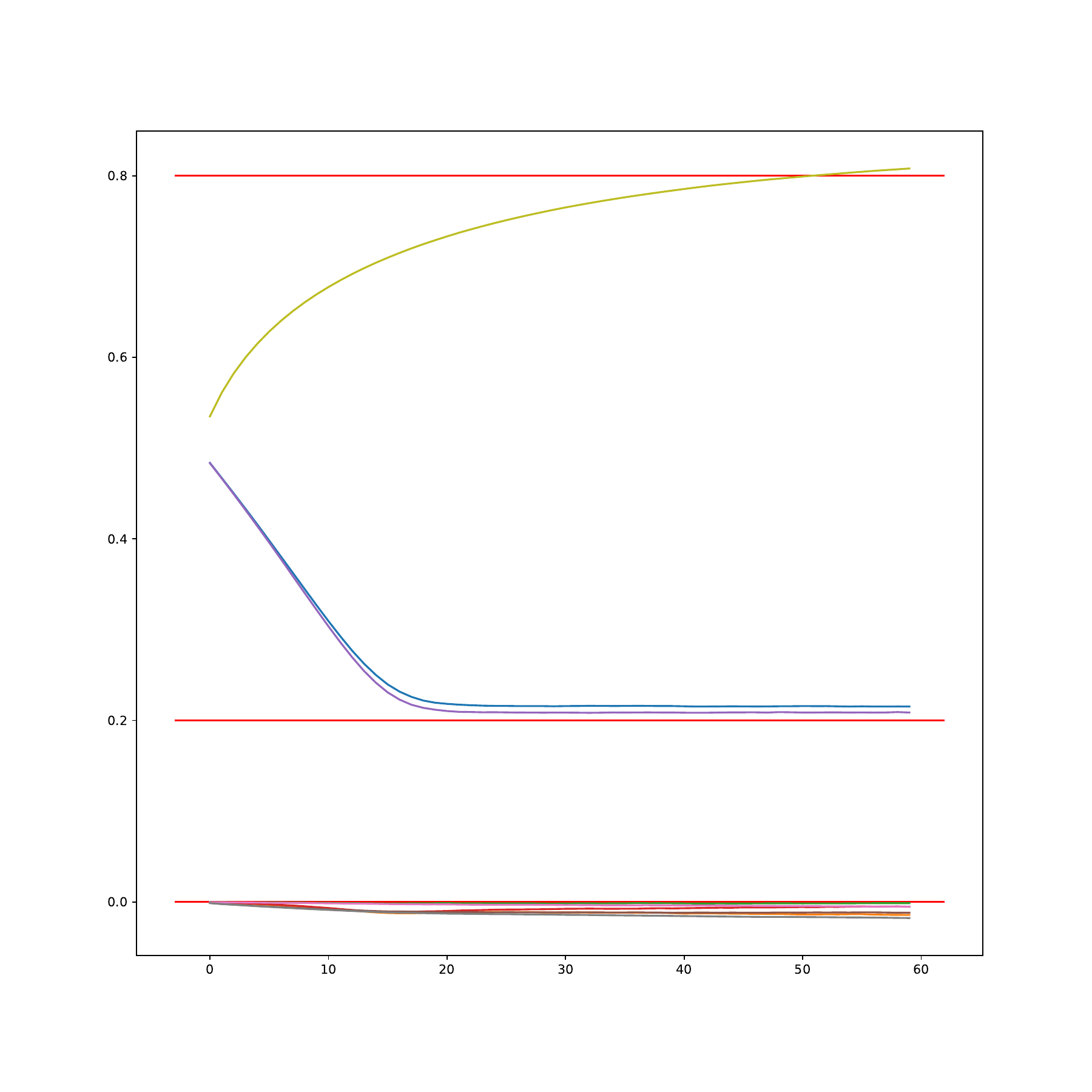}
  }
   \hfill
    \subfloat{
    \includegraphics[width=.40\linewidth,clip=true,trim=50 50 50 50]{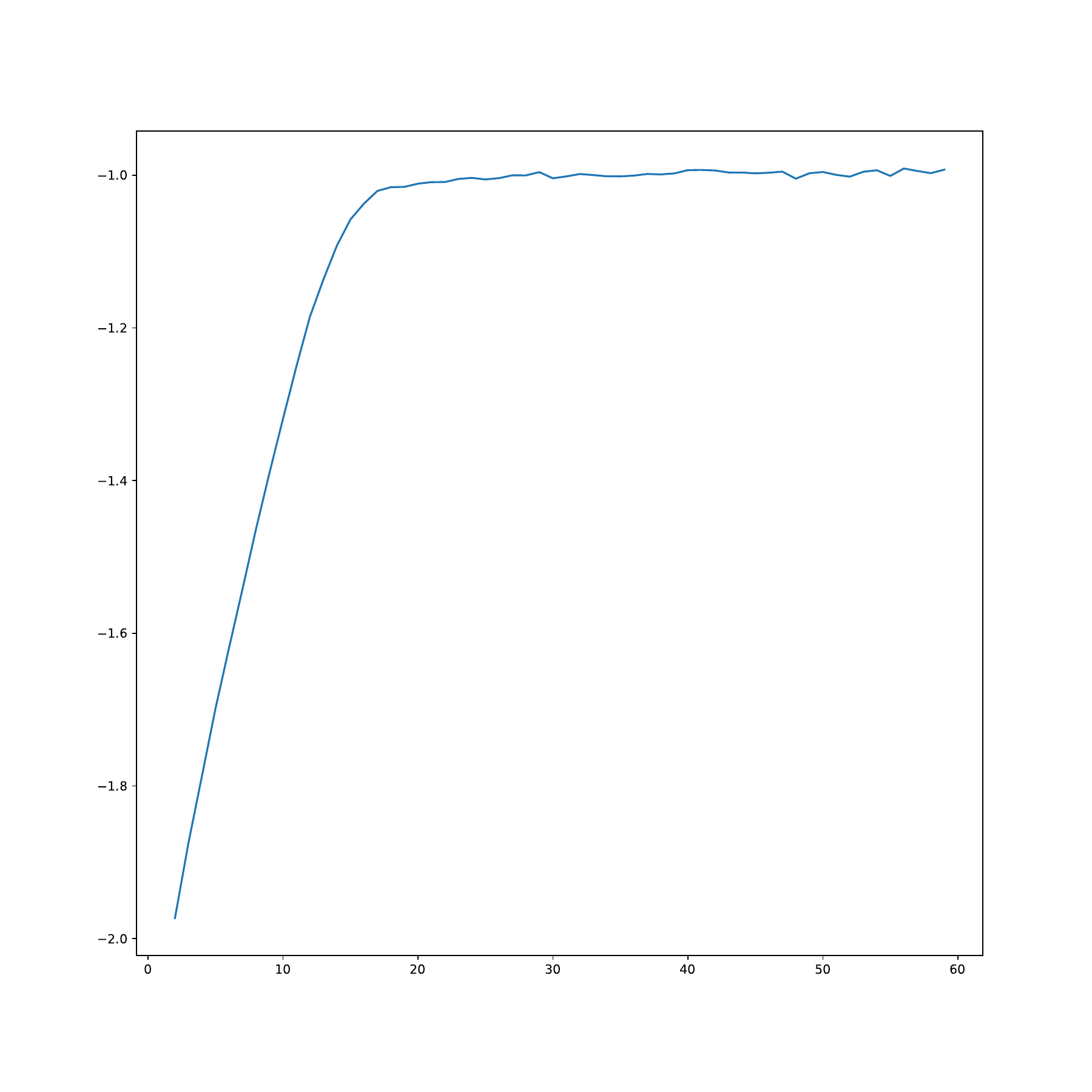}
    }
    \caption{The importance sampling technique applies to estimate the metric on the Lie group $\SO(3)$. Sampling a Brownian motion from an underlying unknown metric, we obtain convergence to the true underlying metric using an iterative MLE method. Here we sampled $4$ guided bridges per observation, providing a relatively smooth iterative likelihood. (Top left) Estimation of the unknown underlying metric using bridge sampling, starting from the metric $\text{diag}(1, 1, 1)$. Here the true metric is the diagonal matrix $\text{diag}(0.2, 0.2, 0.8)$ represented by the red lines. The diagonal is represented by the colors diag(purple,blue,yellow). (Top right) The correspondong iterative log-likelihood. (Bottom left) Estimation of the unknown underlying metric using bridge sampling, starting from the metric $\text{diag}(0.5, 0.5, 0.5)$. (Bottom right) The corresponding iterative log-likelihood.}\label{fig: Metric estimation on SO(3)}
\end{figure}

Figure \ref{fig: Metric estimation on SO(3)} illustrates how importance sampling on $\SO(3)$ leads to a metric estimation of the underlying unknown metric, which generated the Brownian motion. We sampled $128$ points as endpoints of a Brownian motion from the metric $\text{diag}(0.2, 0.2, 0.8)$, and used $20$ time steps to sample $4$ bridges per observation. An iterative MLE method using gradient descent with a learning rate of $0.2$ and an initial guess of the metric being $\text{diag}(1, 1, 1)$ yielded a convergence to the true metric. Note that in each iteration, the logarithmic map changes as can be seen from Algorithm~\ref{alg: iterative maximum likelihood}.

\begin{figure}[H]
  \subfloat{
        \includegraphics[width=.12\linewidth,clip=true,trim=200 200 150 200]{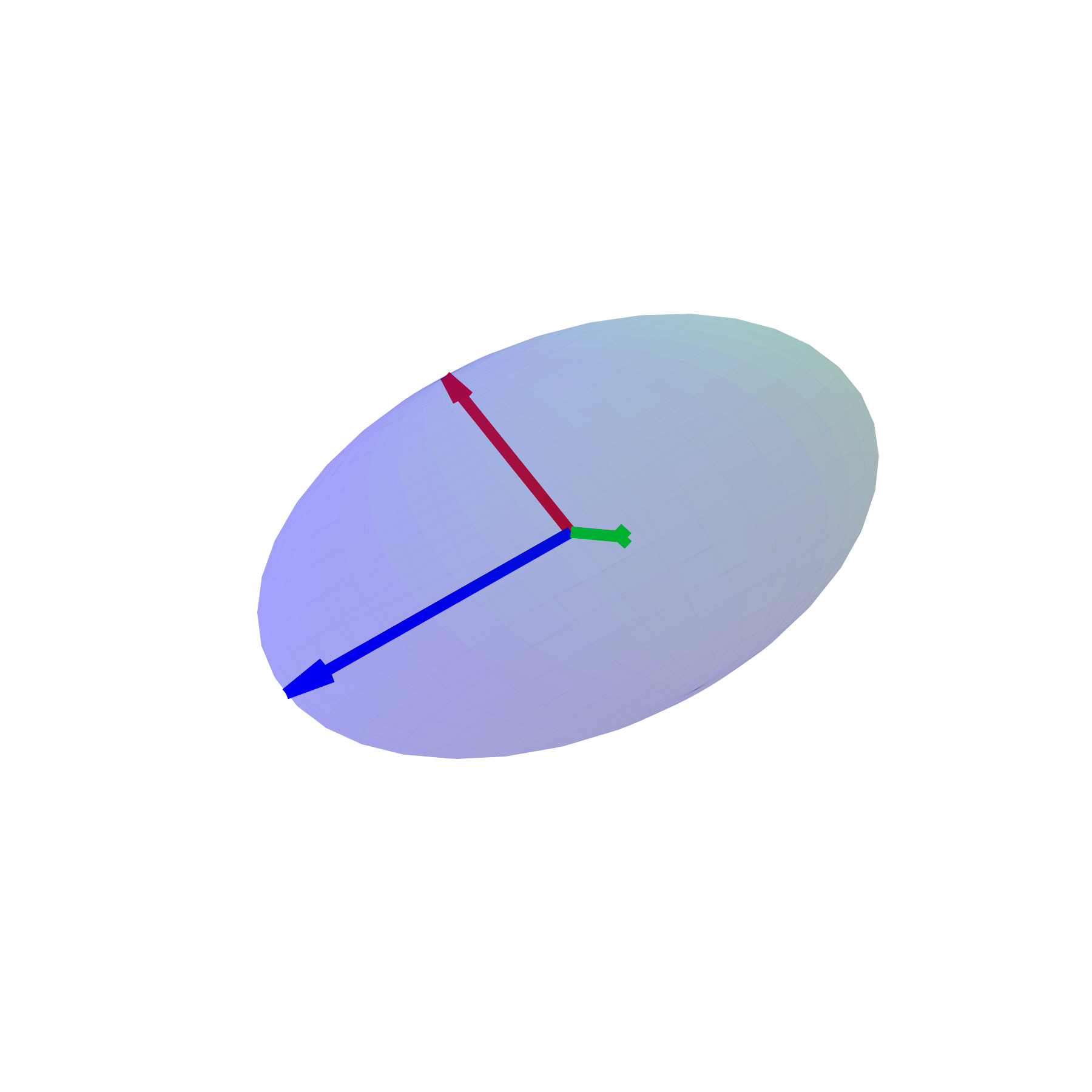} 
      }
    \hfill
    \subfloat{
    \includegraphics[width=.13\linewidth,clip=true,trim=170 200 140 200]{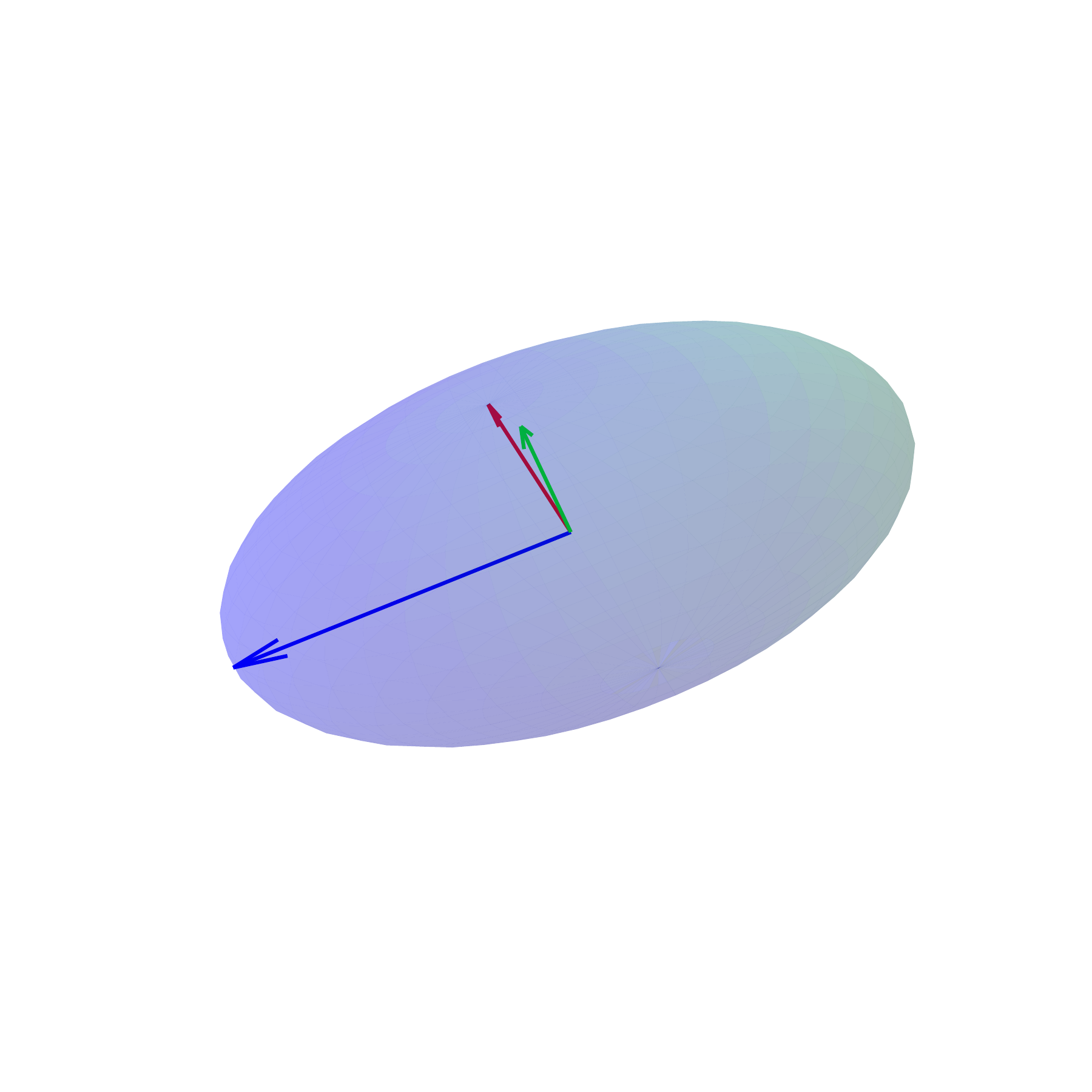}
  }
   \hfill
    \subfloat{
    \includegraphics[width=.13\linewidth,clip=true,trim=200 200 150 200]{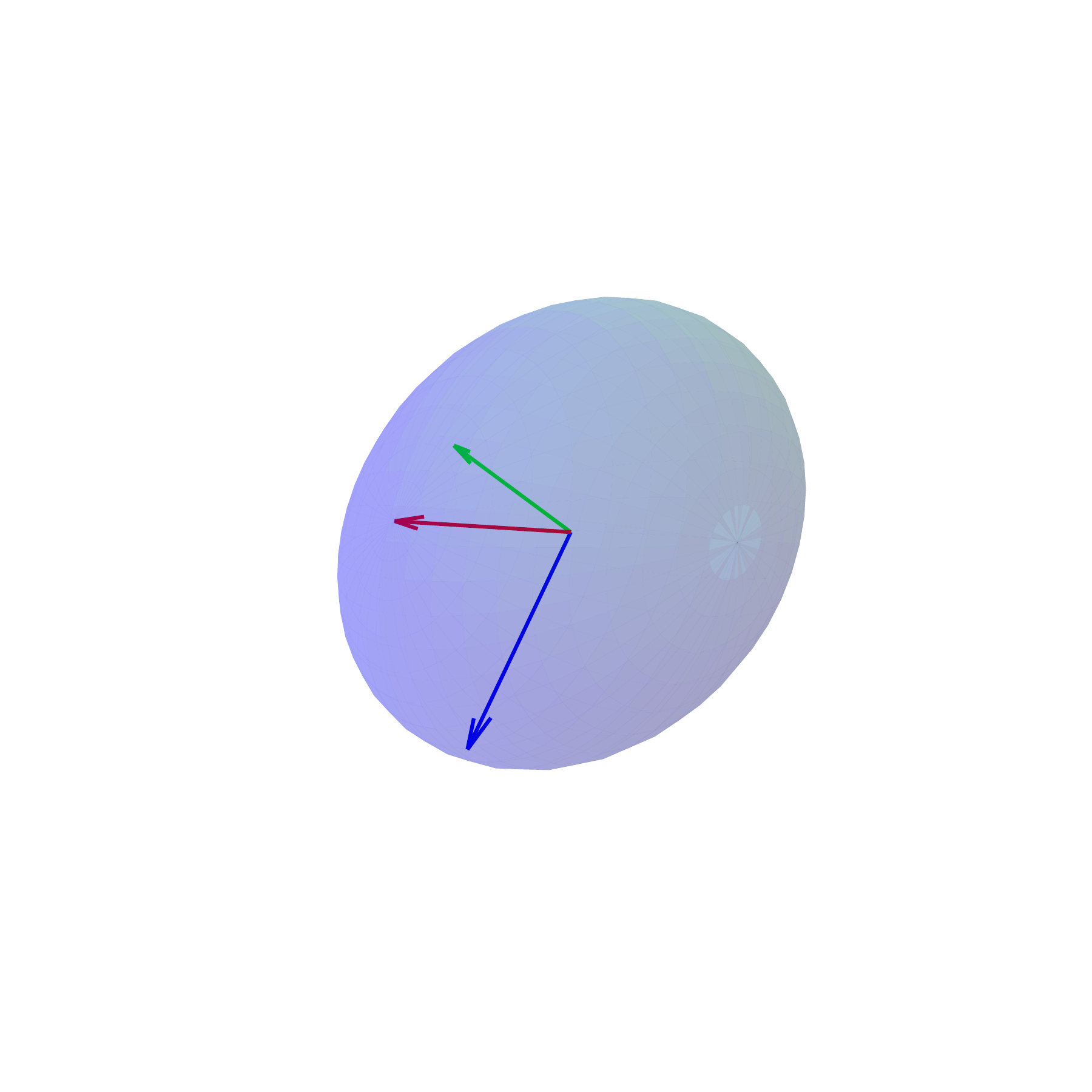}
  }
  \hfill
    \subfloat{
    \includegraphics[width=.14\linewidth,clip=true,trim=200 200 150 200]{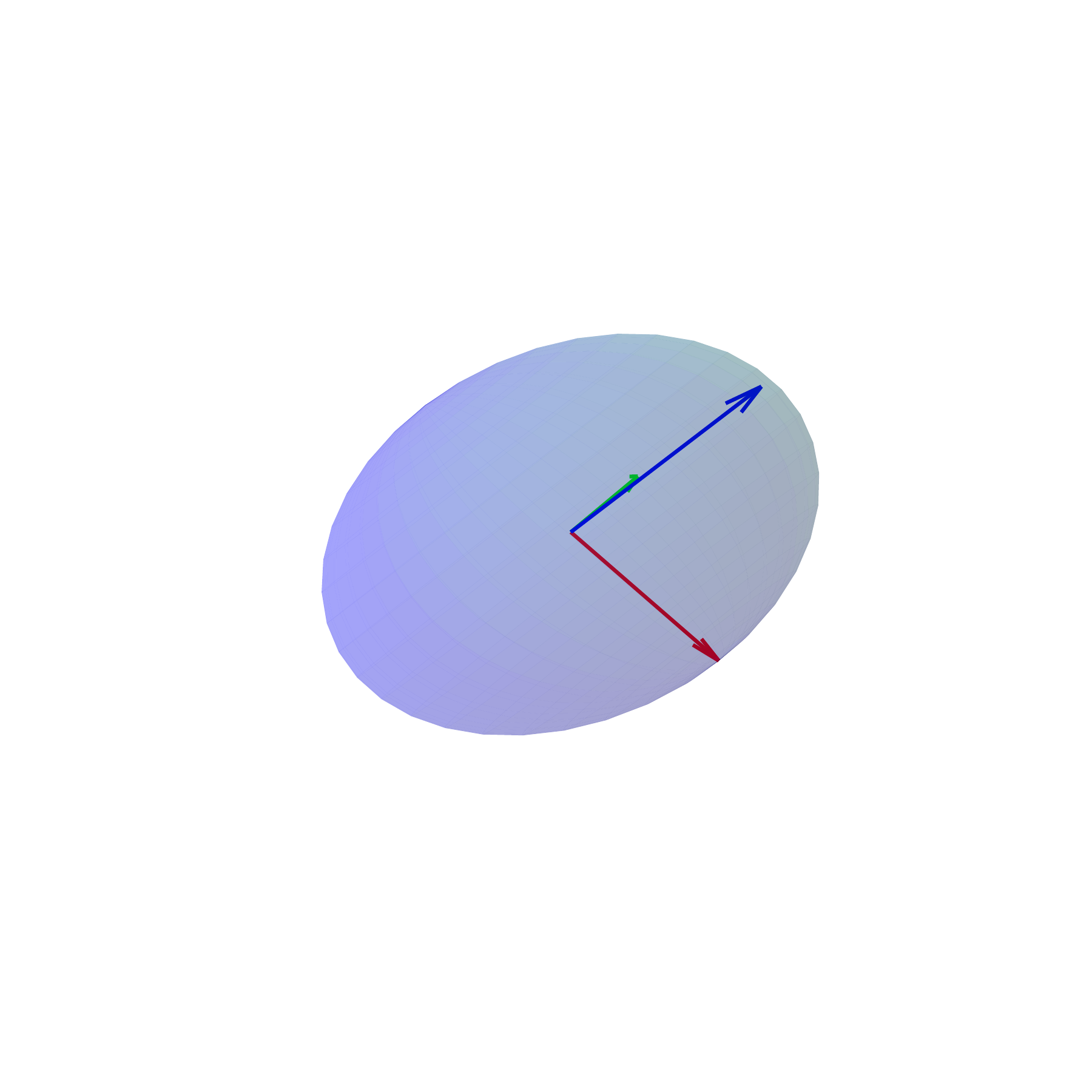}
  }
  \hfill
    \subfloat{
    \includegraphics[width=.13\linewidth,clip=true,trim=200 200 150 200]{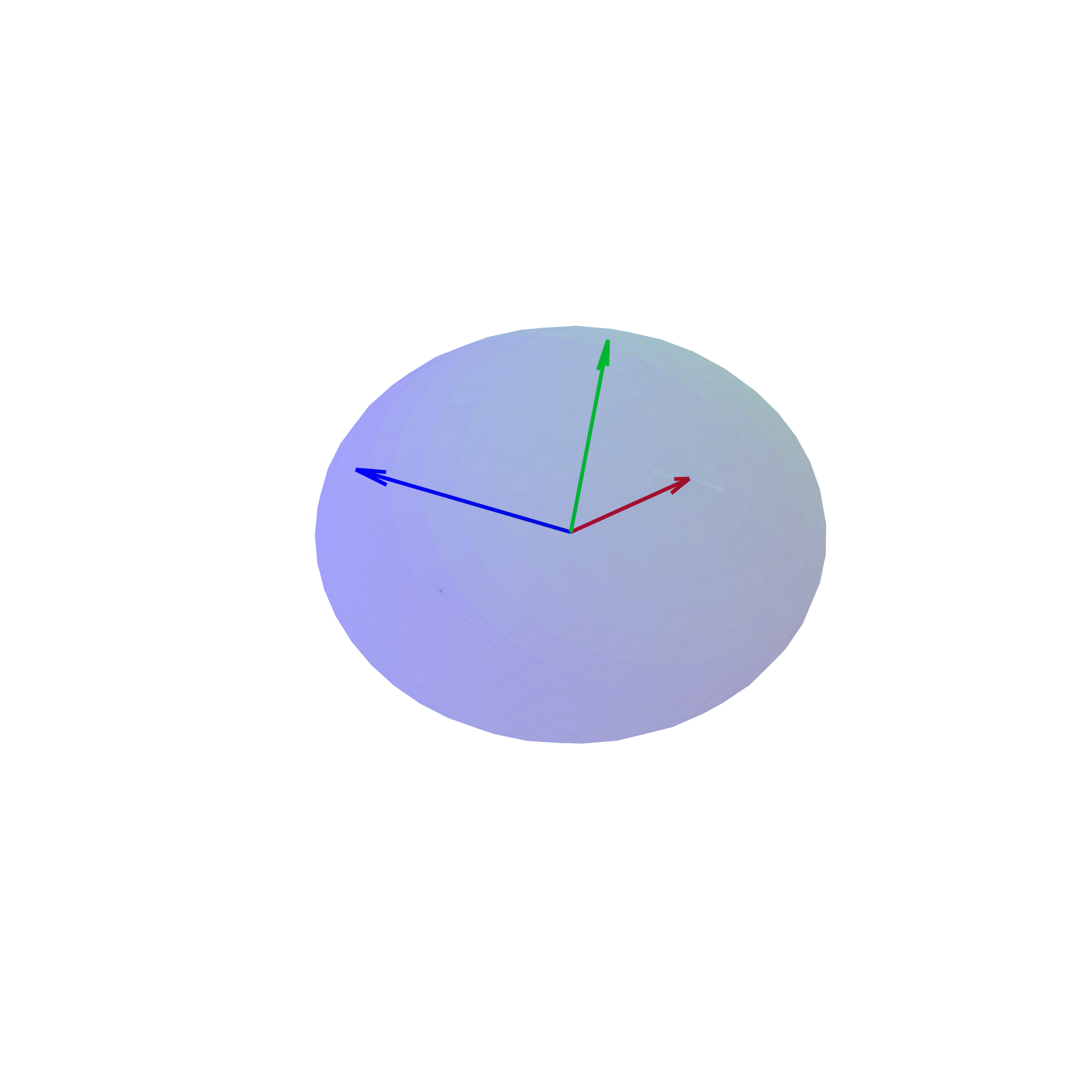}
  }
   \hfill
    \subfloat{
    \includegraphics[width=.13\linewidth,clip=true,trim=200 200 150 200]{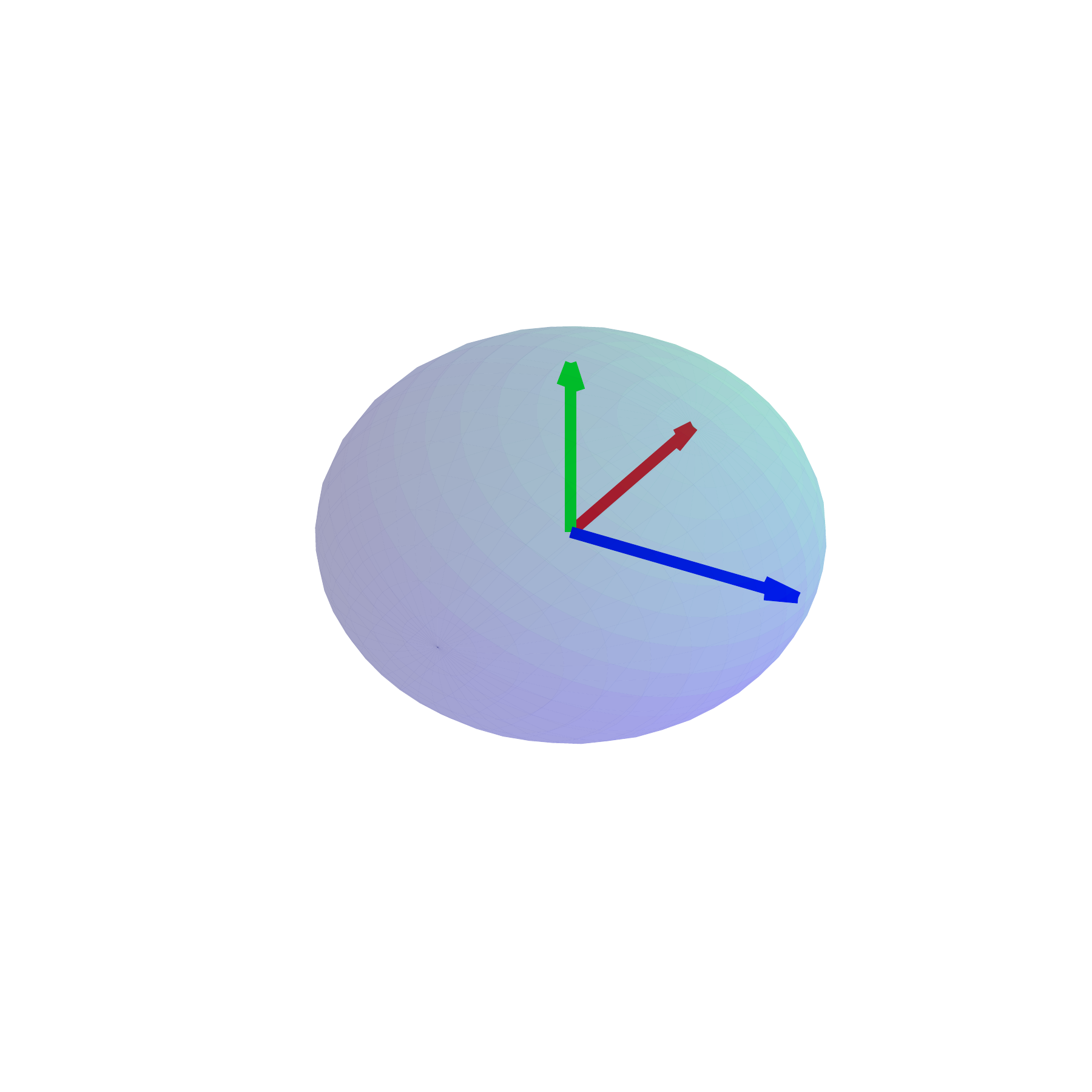}
    }
    \newline
    \subfloat{
        \includegraphics[width=.12\linewidth,clip=true,trim=200 200 150 200]{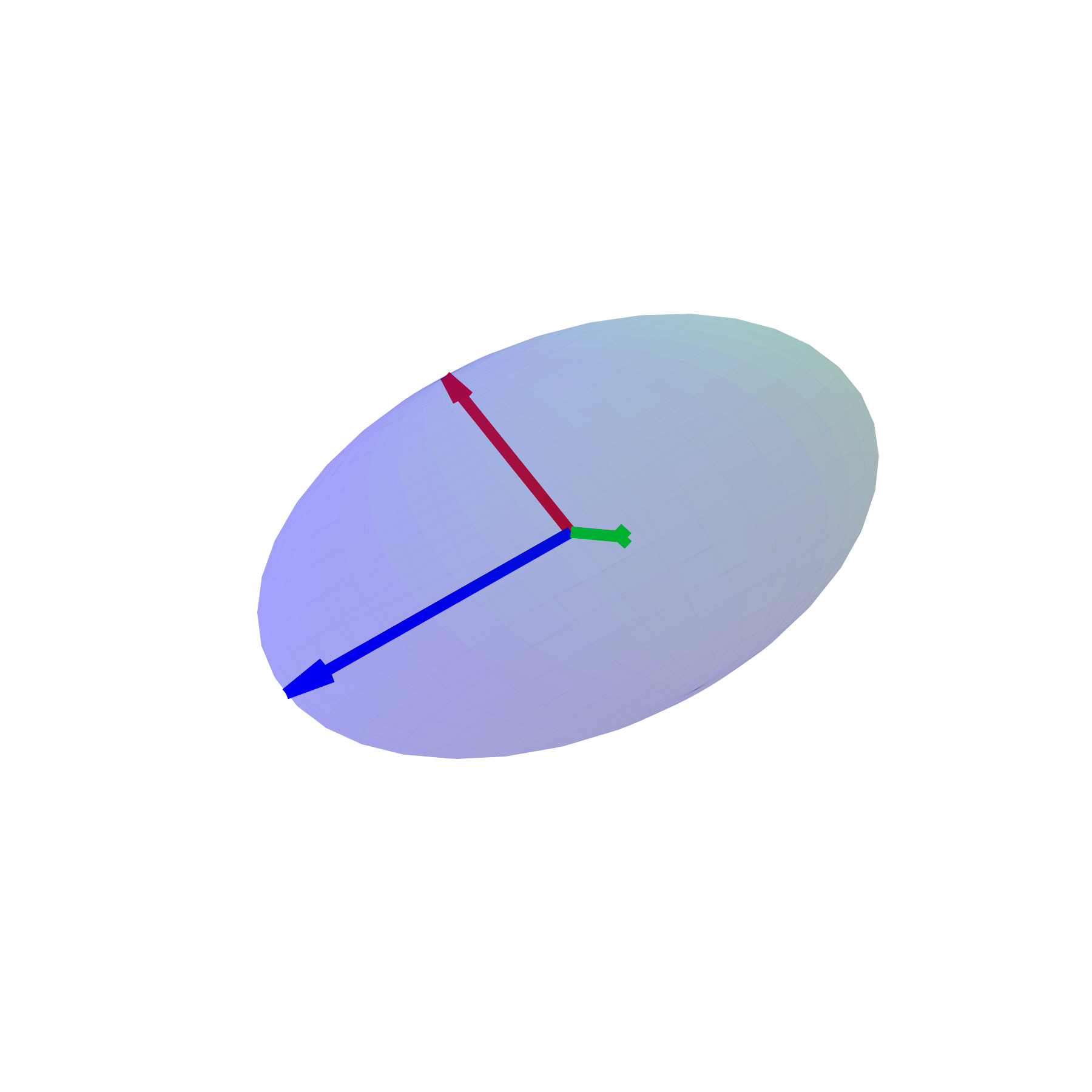} 
      }
    \hfill
    \subfloat{
    \includegraphics[width=.13\linewidth,clip=true,trim=200 200 150 200]{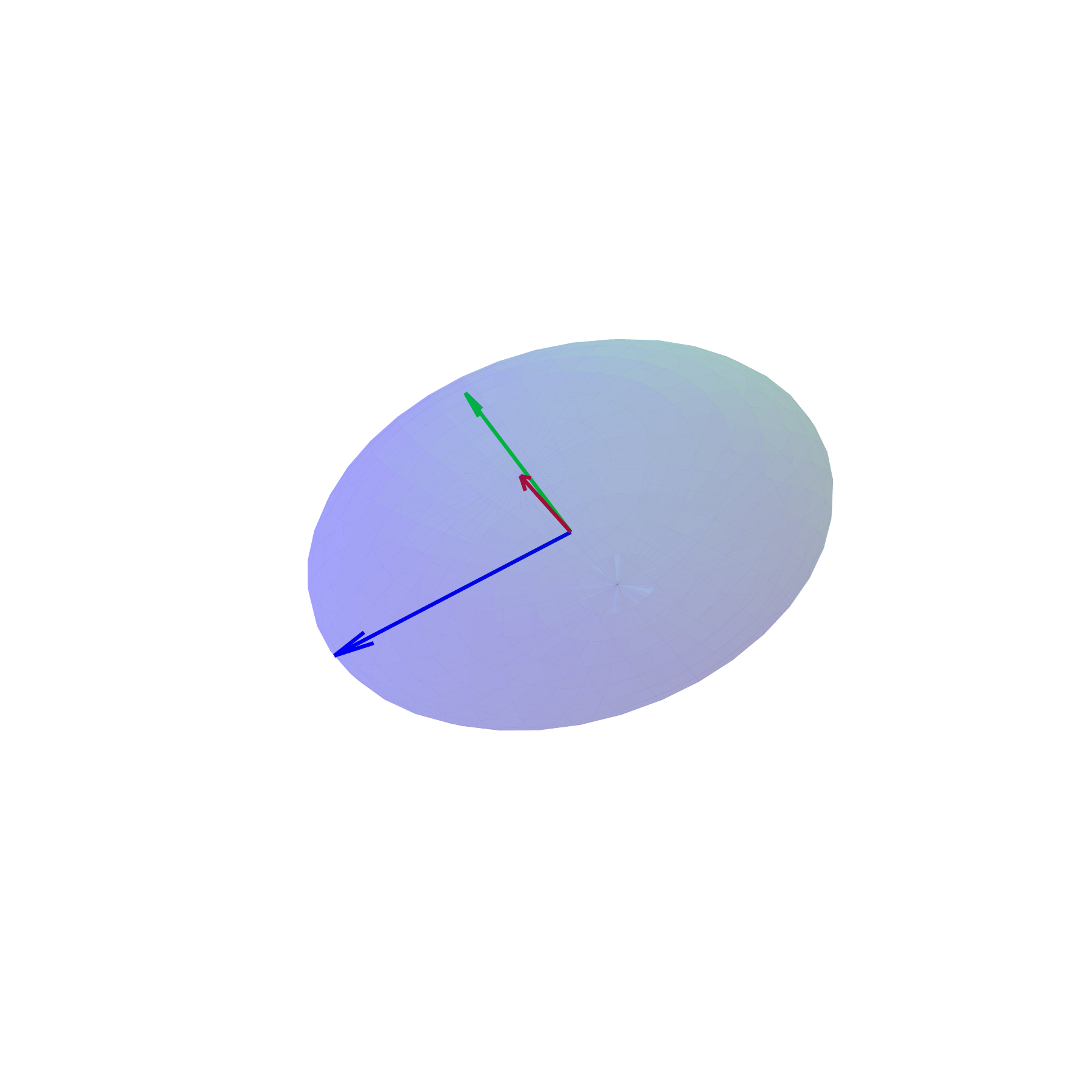}
  }
   \hfill
    \subfloat{
    \includegraphics[width=.13\linewidth,clip=true,trim=200 200 150 200]{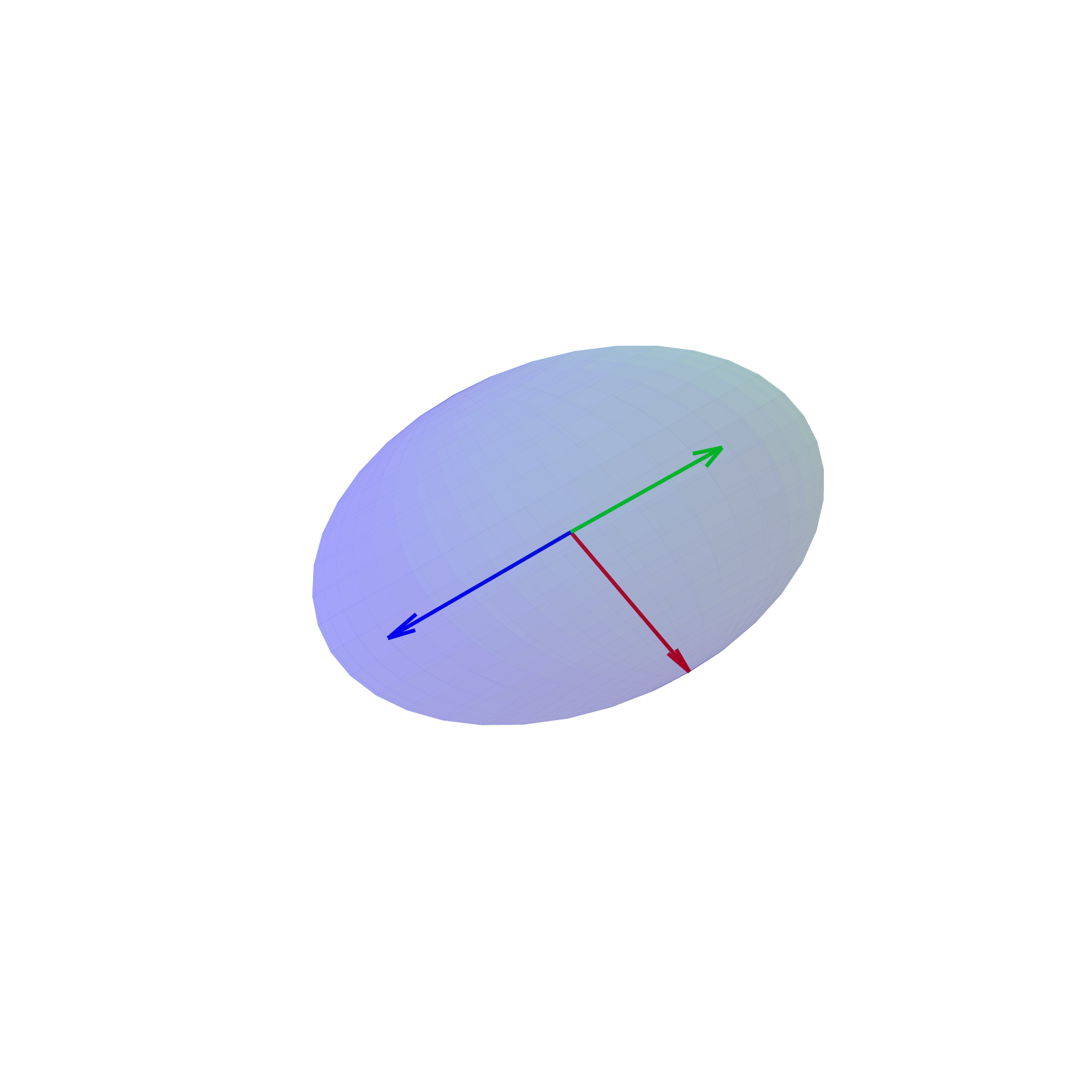}
  }
  \hfill
    \subfloat{
    \includegraphics[width=.13\linewidth,clip=true,trim=200 200 150 200]{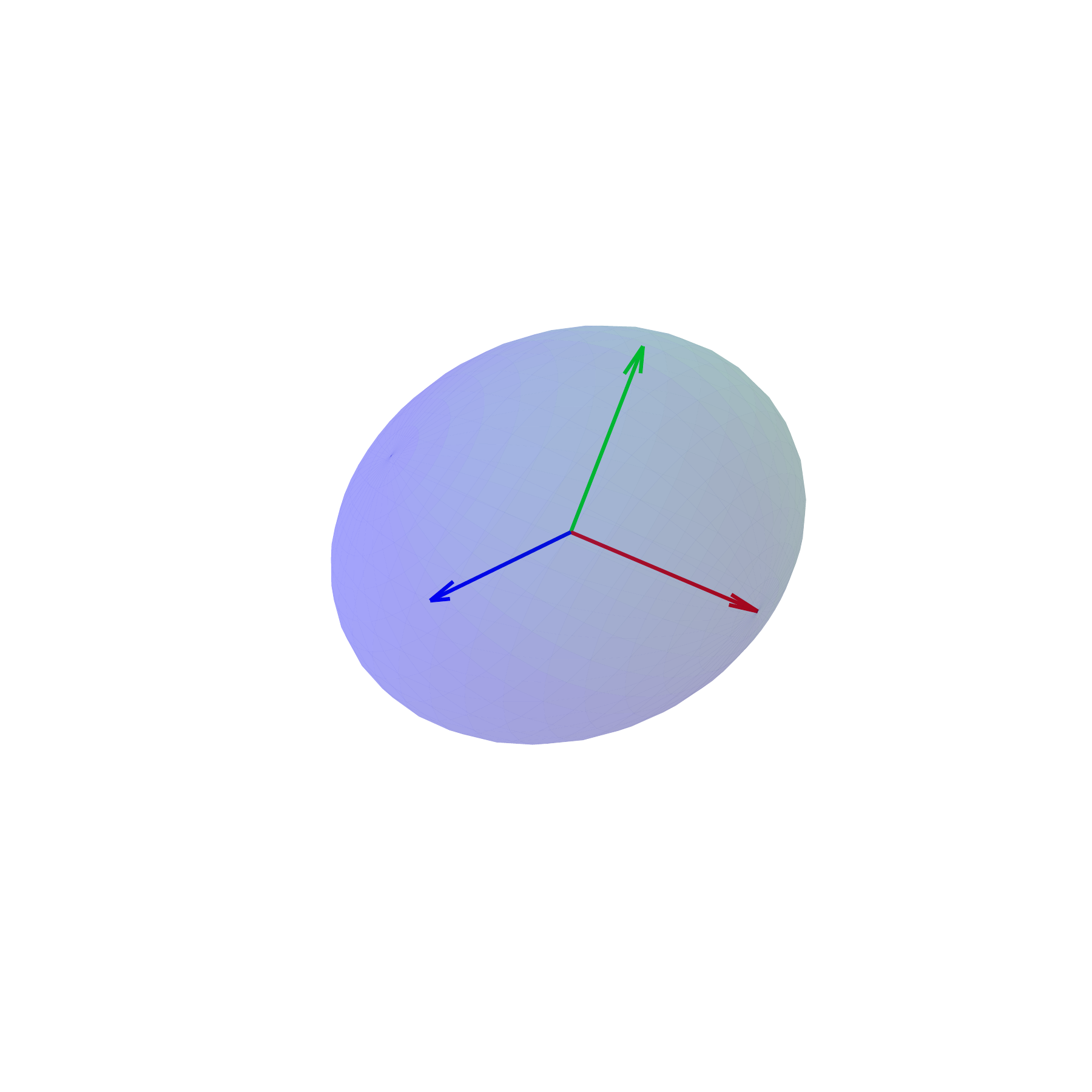}
  }
  \hfill
    \subfloat{
    \includegraphics[width=.13\linewidth,clip=true,trim=200 200 150 200]{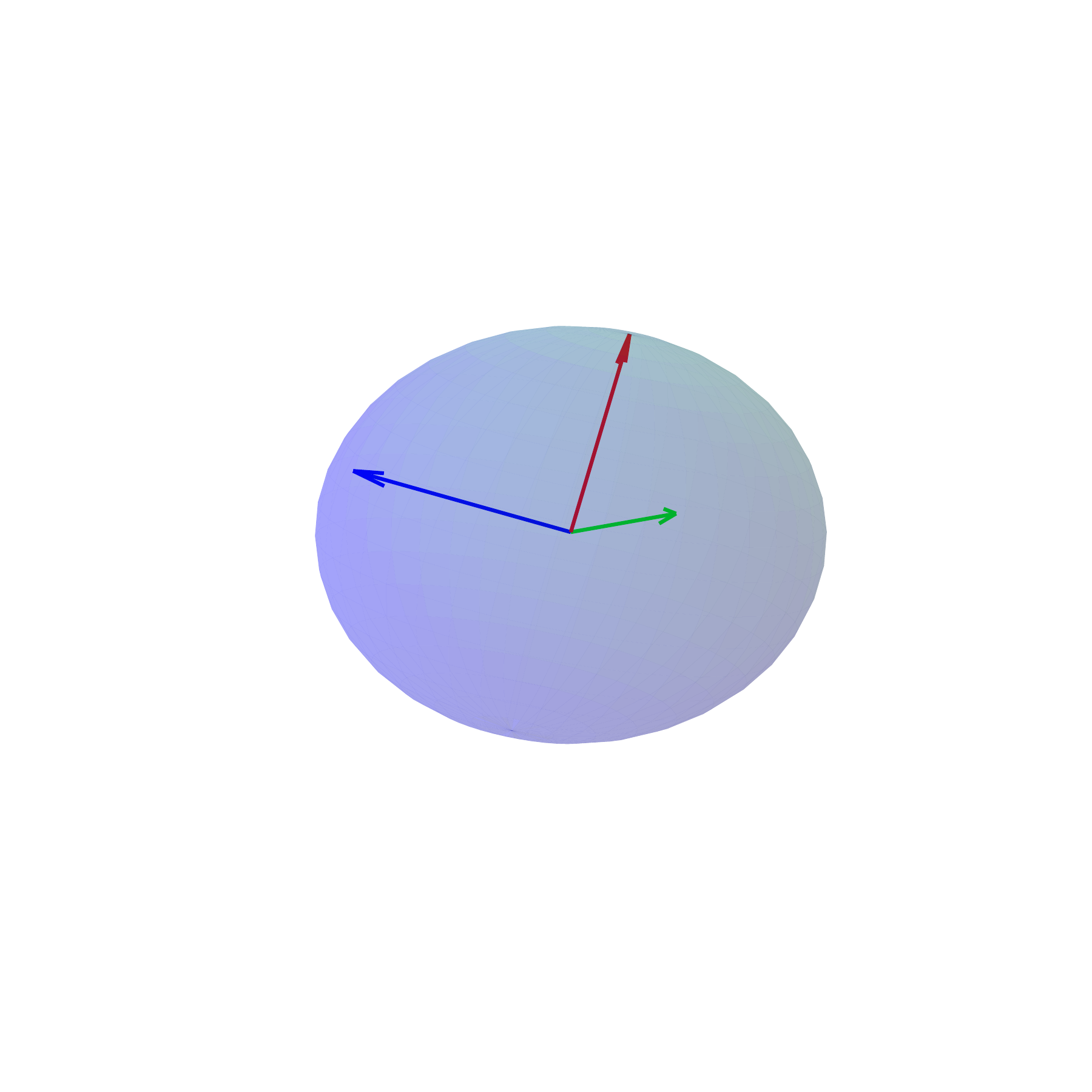}
  }
   \hfill
    \subfloat{
    \includegraphics[width=.13\linewidth,clip=true,trim=200 200 150 200]{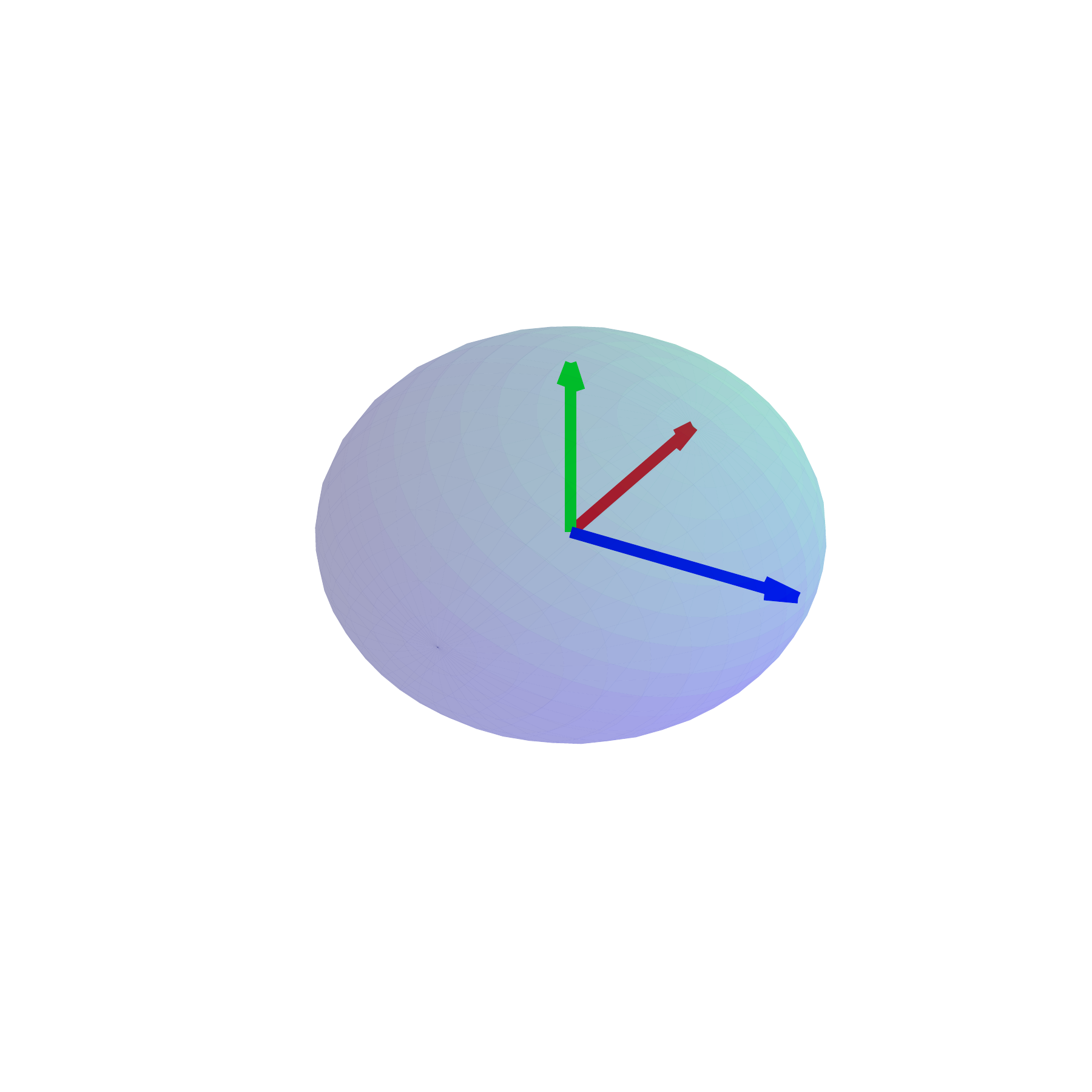}
    }
    \newline
    \subfloat{
        \includegraphics[width=.12\linewidth,clip=true,trim=200 200 150 200]{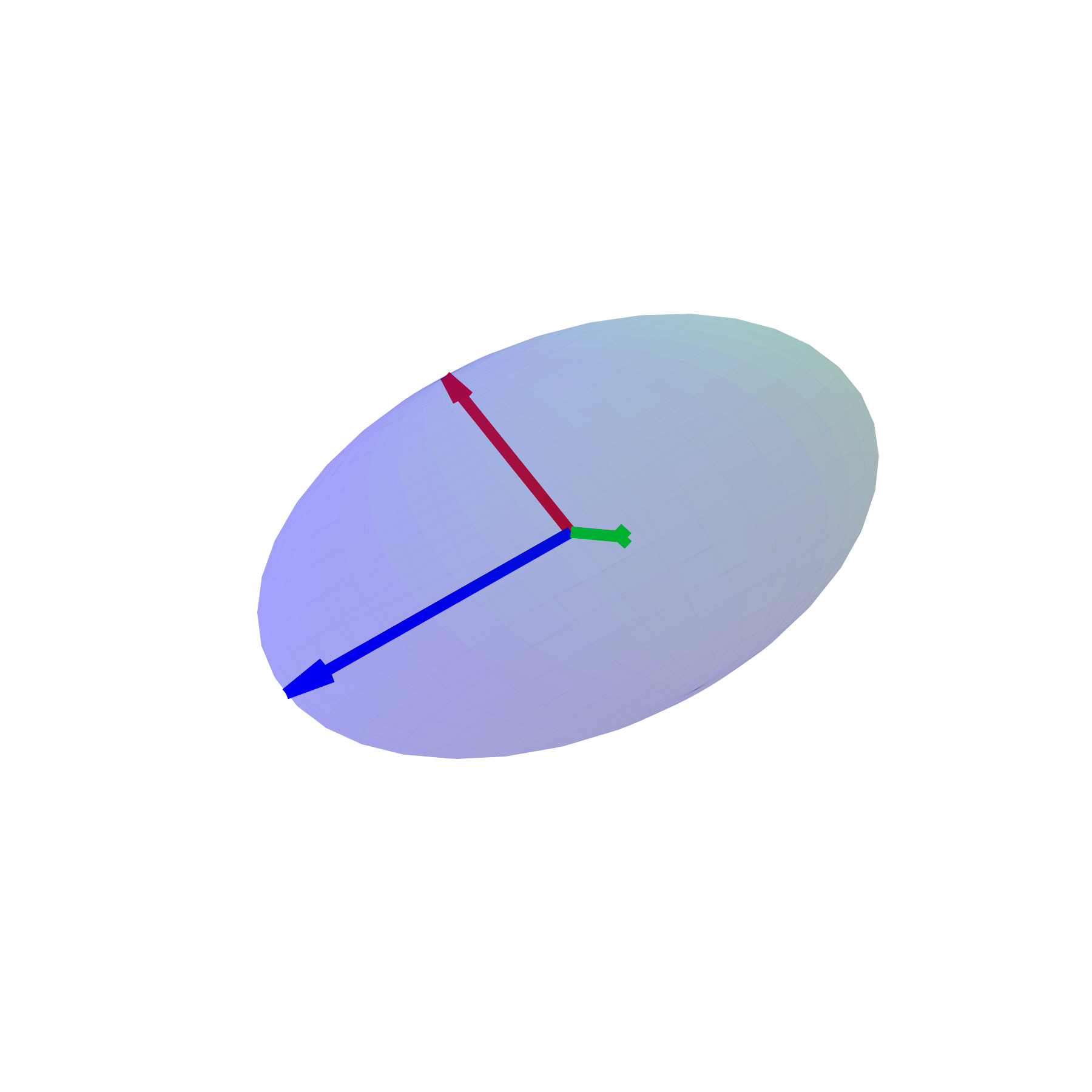} 
      }
    \hfill
    \subfloat{
    \includegraphics[width=.12\linewidth,clip=true,trim=200 200 150 200]{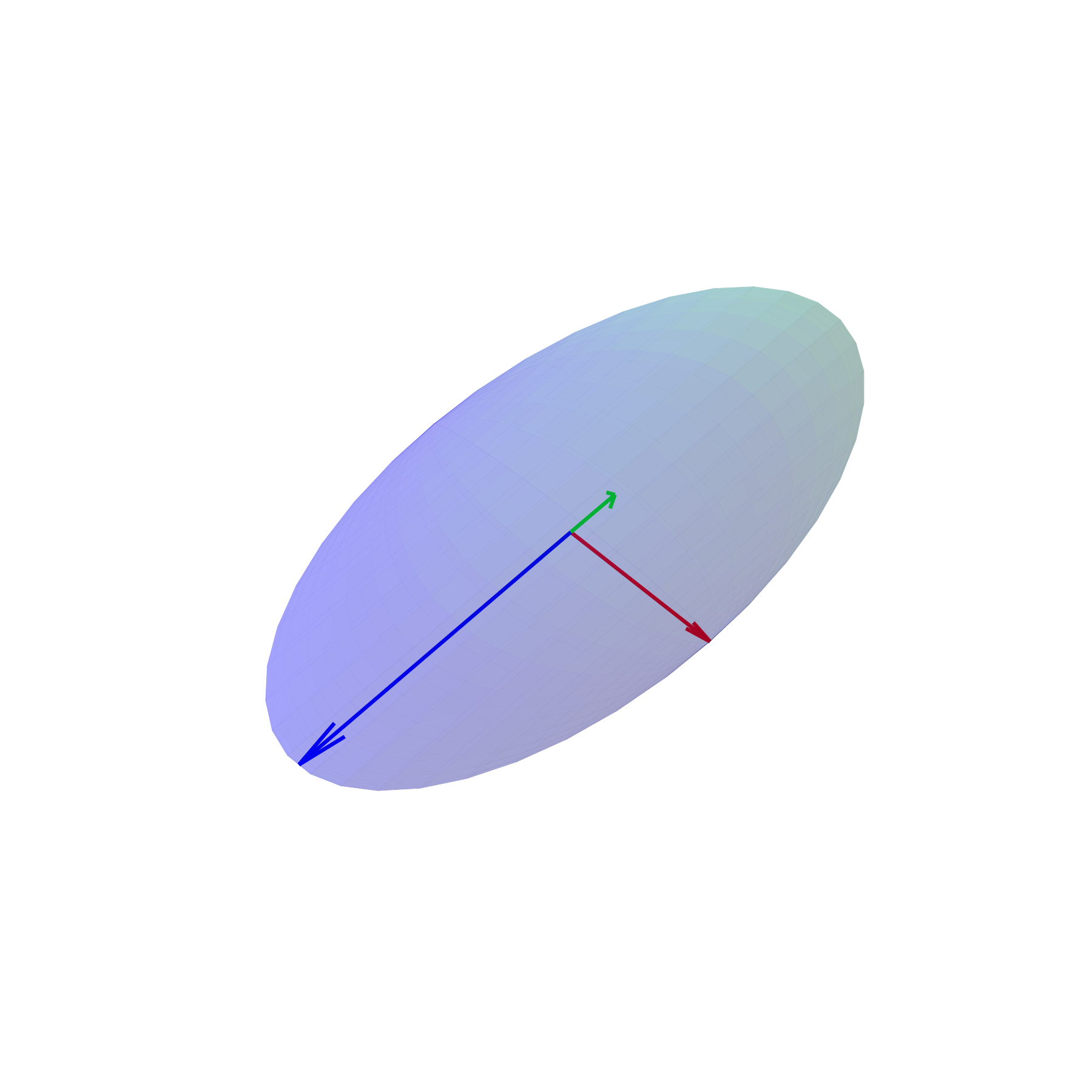}
  }
   \hfill
    \subfloat{
    \includegraphics[width=.13\linewidth,clip=true,trim=200 200 150 150]{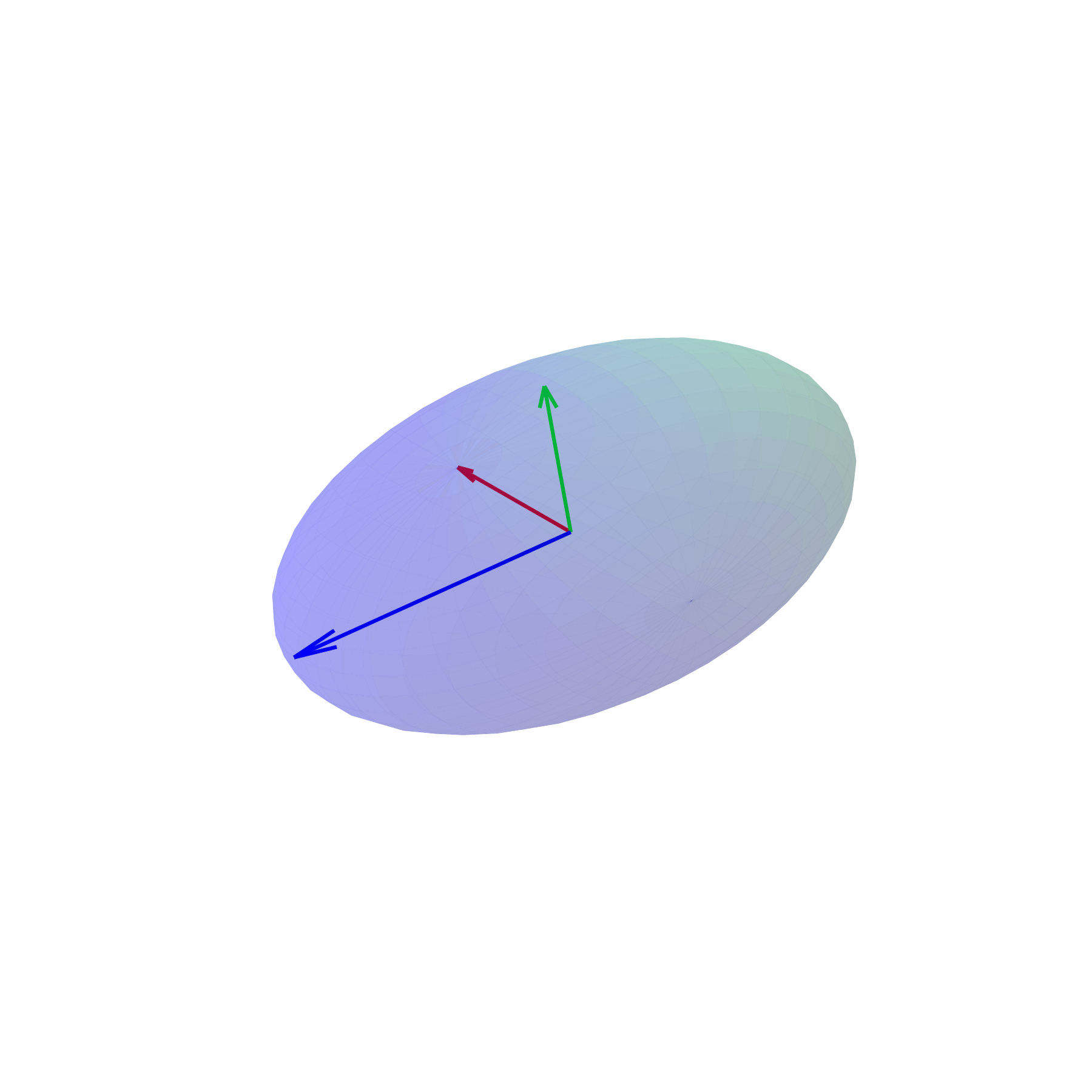}
  }
  \hfill
    \subfloat{
    \includegraphics[width=.13\linewidth,clip=true,trim=200 200 150 150]{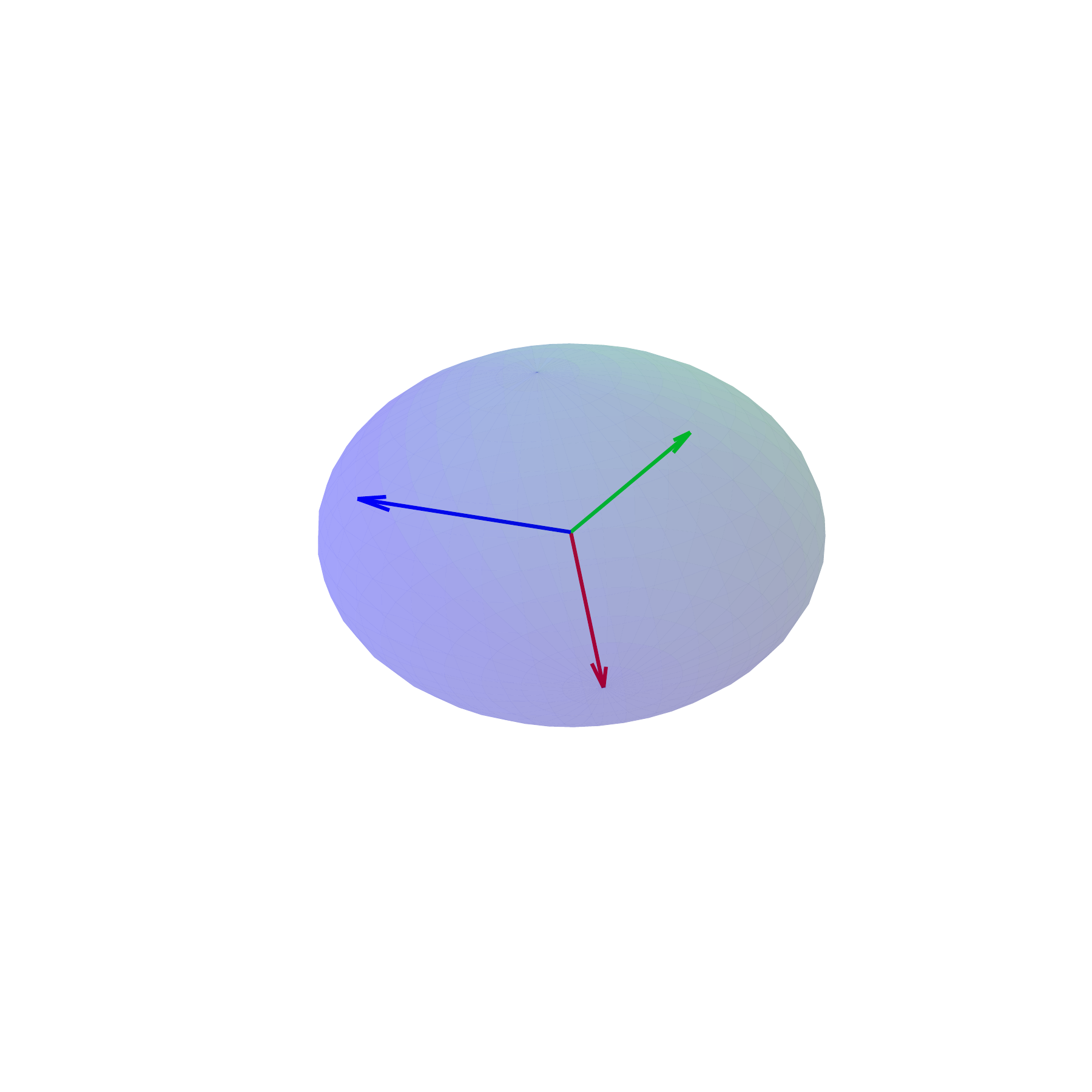}
  }
  \hfill
    \subfloat{
    \includegraphics[width=.13\linewidth,clip=true,trim=200 200 150 200]{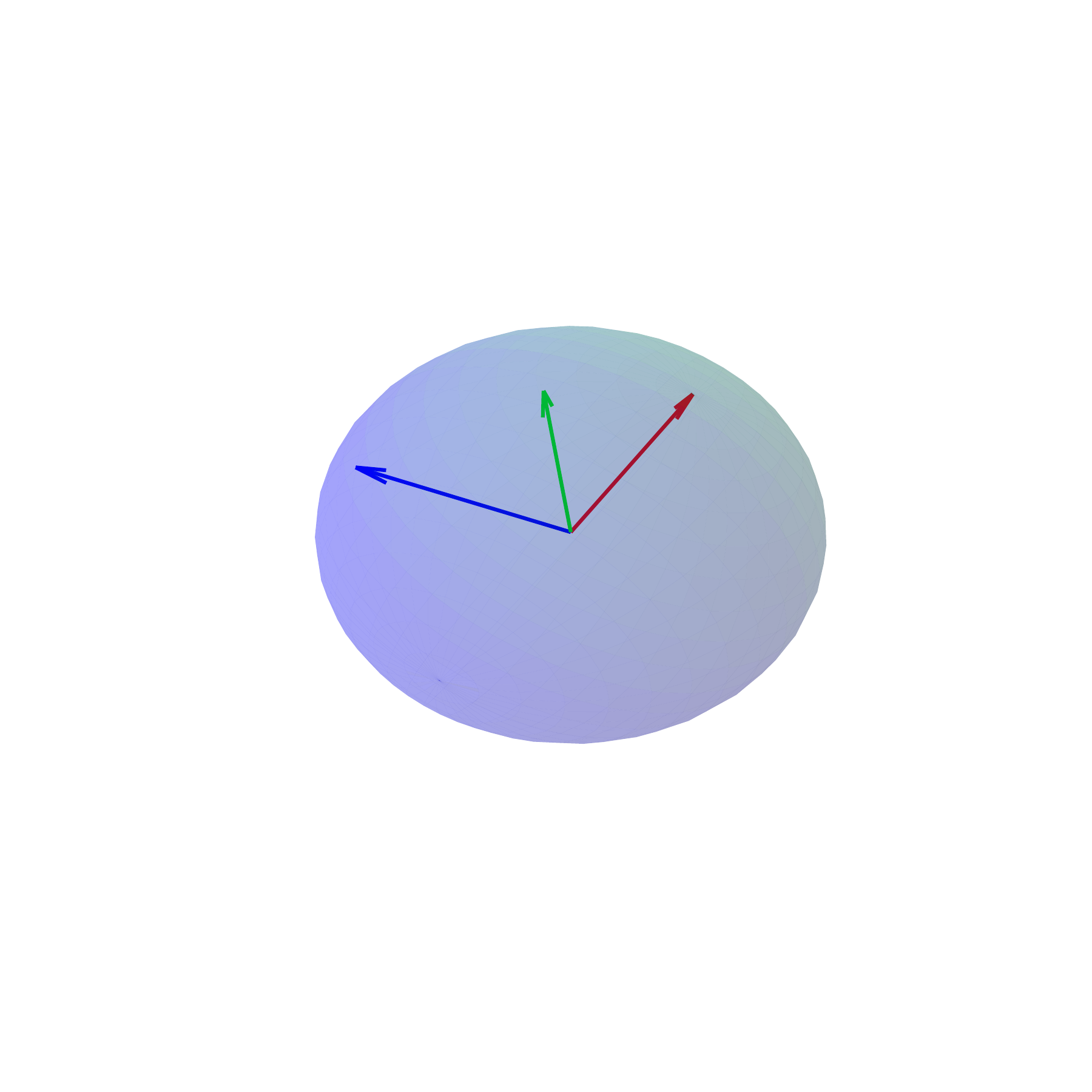}
  }
   \hfill
    \subfloat{
    \includegraphics[width=.13\linewidth,clip=true,trim=200 200 150 200]{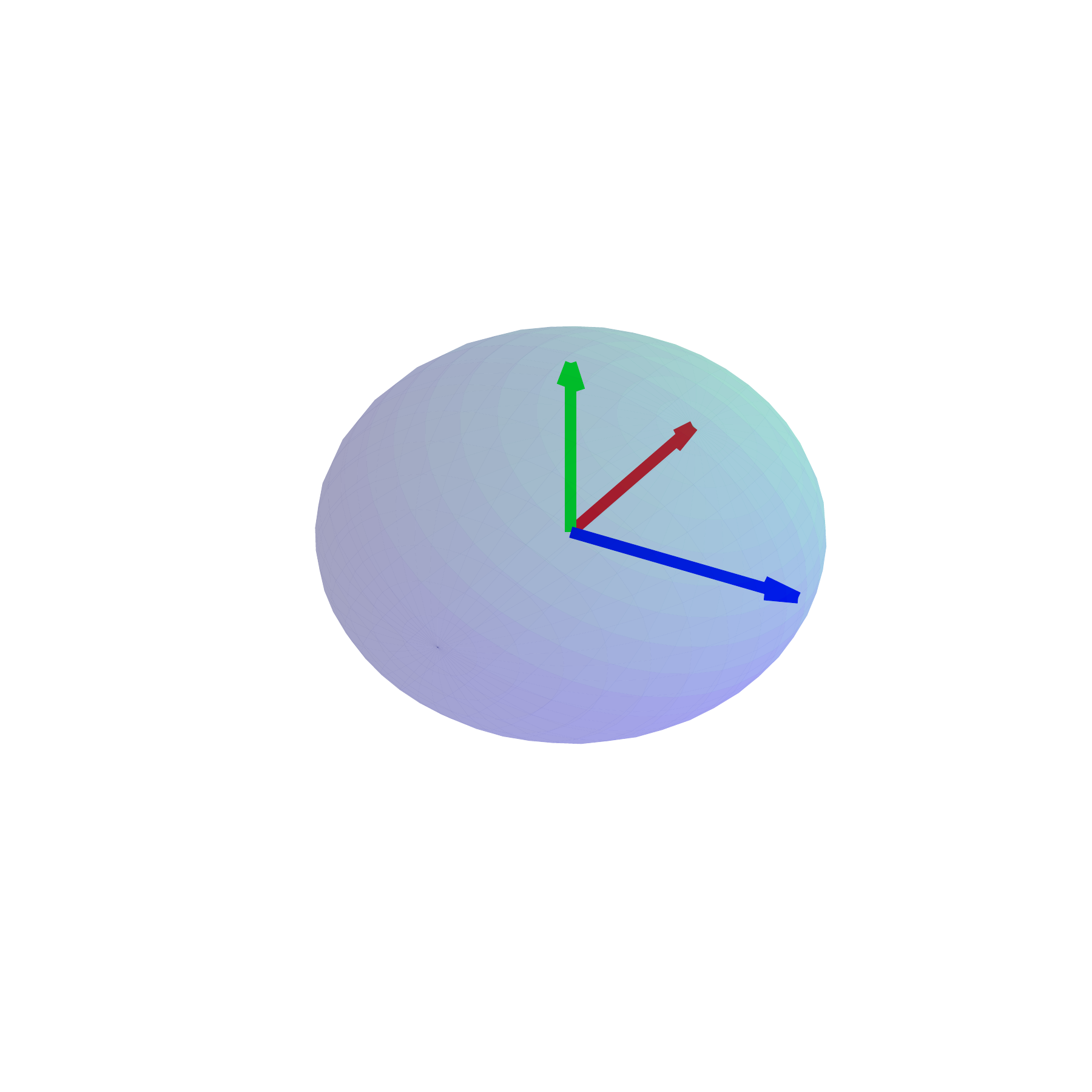}
    }
    \caption{Discrete time observations from three sample paths on $\SPD(3)$. The sample paths are obtained as the pushforward of the Fermi bridge in $\GL_+(3)$. The start and endpoint are the left- and rightmost figures, where the SPD matrices are indicated by the bold face arrows.}
    \label{fig: guided bridge on spd3}
\end{figure}

\subsection{Diffusion-mean estimation on the space of symmetric positive definite matrices}\label{sec: bridge sampling on spd3}

The space of symmetric positive definite ($\SPD$) matrices is used in in a range of applied fields, one example being diffusion tensor imaging where element of $\SPD(3)$ models anisotropic diffusion of water molecules in each position of the imaged domain. The $\SPD$ matrices constitute a smooth incomplete manifold when endowed with the Euclidean metric of matrices \cite{pennec2019riemannian}. However, endowing the space of $\SPD$ matrices with either the Log Euclidean or the Affine invariant metric makes the space geodesically complete, i.e., the exponential map is a global diffeomorphism. The space of $\SPD(3)$ matrices can be regarded as the homogeneous space $\GL_+(3)/\SO(3)$ of invertible matrices with positive determinants being rotationally invariant to three-dimensional rotations.
Figure~\ref{fig: guided bridge on spd3} illustrates the discrete time observations from three different sample paths in $\SPD(3)$ arising from the pushforward of a Fermi bridge in $\GL_+(3)$.

\begin{figure}[H]
    \subfloat{
    \includegraphics[width=.45\linewidth,clip=true,trim=50 50 50 50]{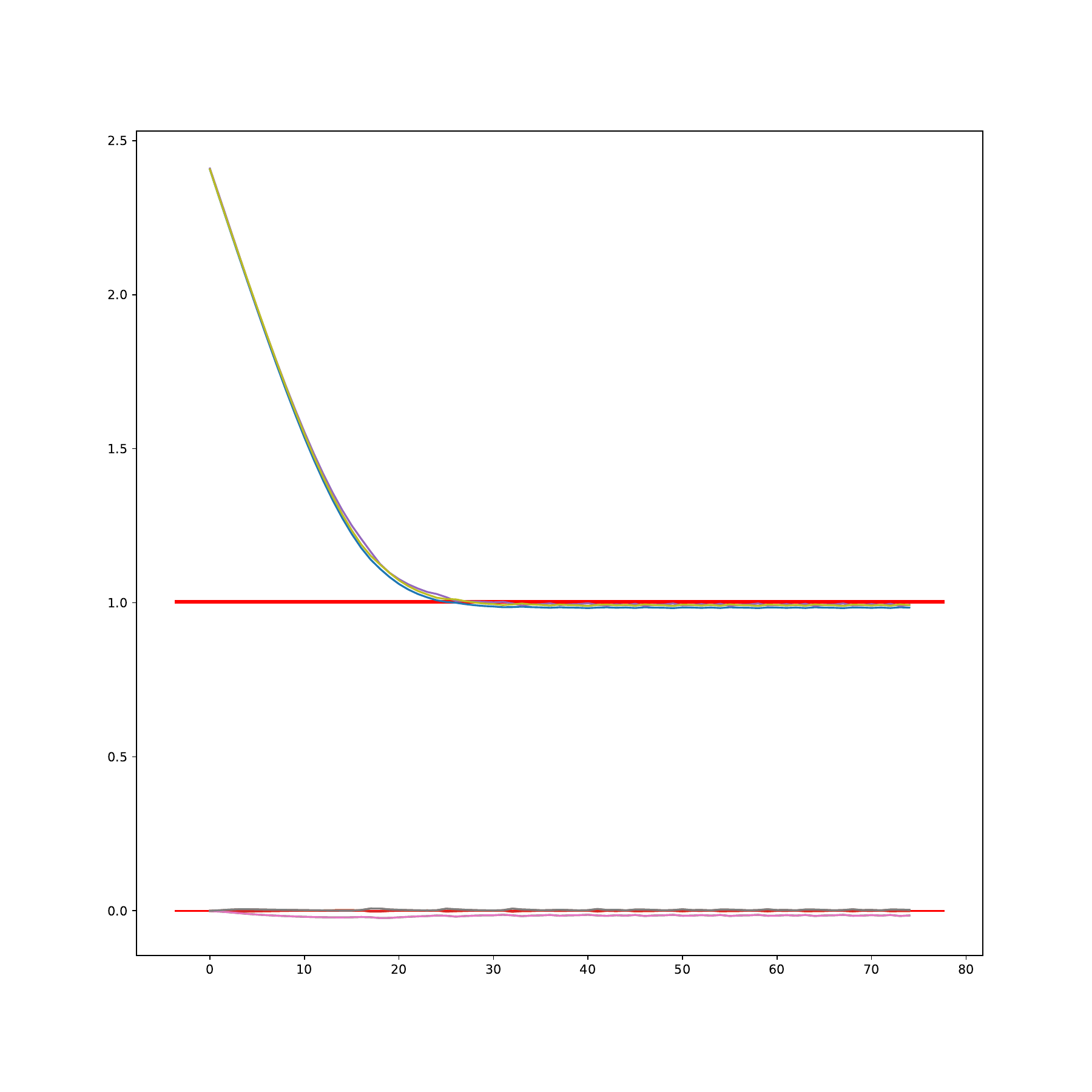}
  }
   \hfill
    \subfloat{
    \includegraphics[width=.45\linewidth,clip=true,trim=50 50 50 50]{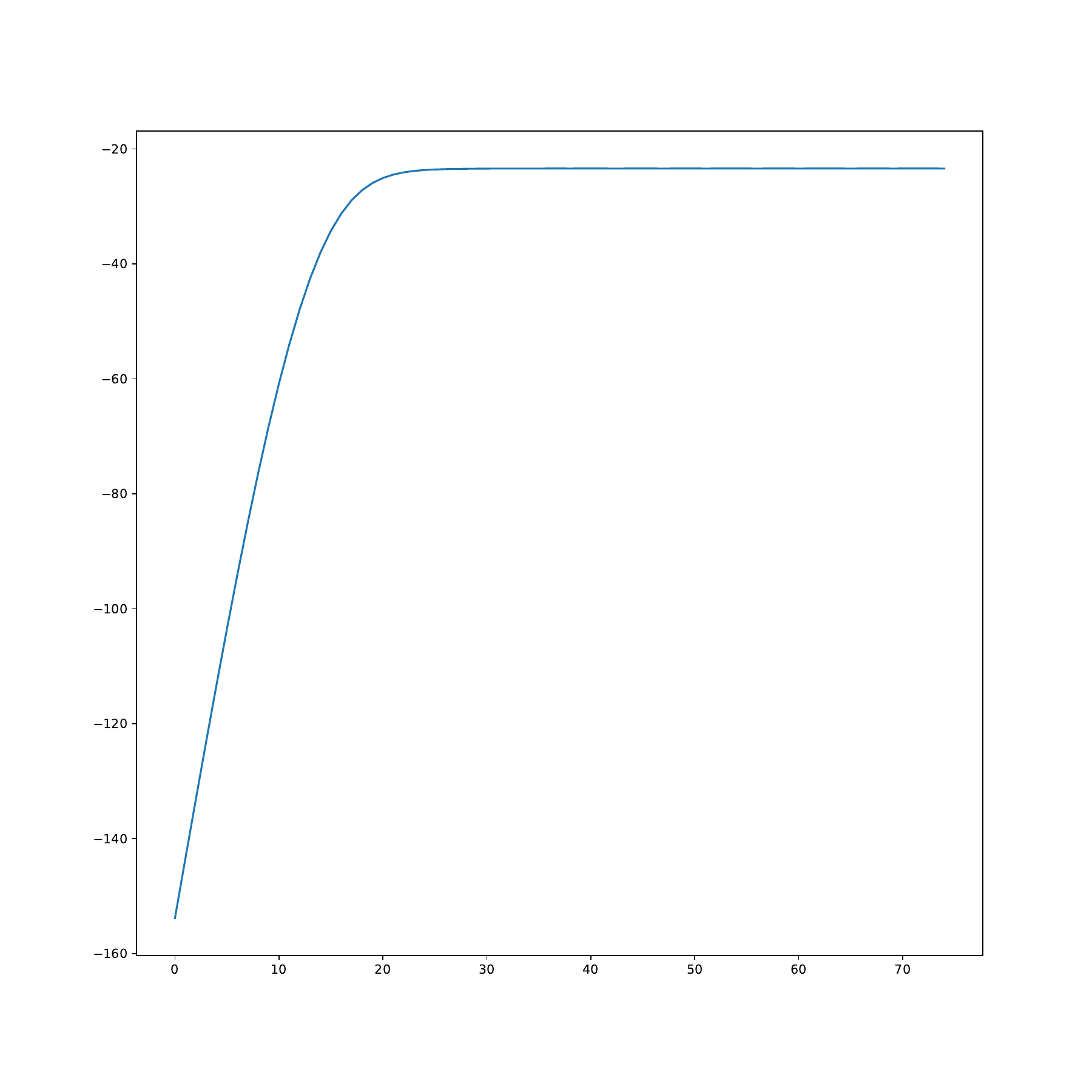}
    }
    \caption{Given $256$ data points in $\SPD(3)$, we estimated the diffusion mean on the homogeneous space by sampling bridges in the top space conditioned on the fibers. The iterative MLE in Algorithm~\ref{alg: iterative maximum likelihood} yielded the convergence of the diffusion mean parameter, using a learning rate of $0.005$ and one bridge sample per observation. (Left) The purple, blue, and yellow line correspond to the diagonal, while the remaining colors represent the off diagonal. The true mean value is the identity matrix indicated by the red lines. (Right) The corresponding iterative log-likelihood.
    }\label{fig: diffusion mean estimation on gl3 and spd3}
\end{figure}

In Figure~\ref{fig: diffusion mean estimation on gl3 and spd3}, the bridge sampling scheme derived above is used to obtain an estimate of the diffusion-mean \cite{hansen2021diffusion,hansen2021diffusiongeometric} on $\SPD(3)$, by sampling guided bridge processes in the space of invertible matrices with positive determinants $\GL_+(3)$. This sampling method provides an estimate of the density on $\GL_+(3)$, which projects to a density in $\SPD(3)$. Exploiting the resulting density in $\SPD(3)$, an iterative MLE method then yields convergence to the diffusion mean.

\subsection{Density estimation on the two-sphere}\label{sec: density estimation on s2}

As explained in \refsec{sec: guide to nearest point}, we introduced a simulation scheme on specific homogeneous spaces by using guided bridges in the top space conditioned to arrive in the fiber at time $T$. The two-sphere $\mathbb S^2$ can be considered the homogeneous space $\SO(3)/\SO(2)$ of three-dimensional rotations, identifying the subgroup of two-dimensional rotations as a single point. Conditioning on the fiber $\SO(2)$ in $\SO(3)$, we obtain guided bridges on $\mathbb S^2$. In the case of a bi-invariant metric on $G$, the $G$-valued Brownian motion pushes forward to an $M$-valued Brownian motion. The second illustration in Figure~\ref{fig: anisotropic distribution on S2} illustrates the estimated transition density on $\mathbb S^2$ from sampling bridges in the Lie group conditioned on the fiber $\SO(2)$, when the underlying metric is bi-invariant. When altering the metric to a non-invariant variant one, the $G$-Brownian motion does not in general push forward to an $M$-Brownian motion. The non-invariant metrics result in a covariance structure exhibiting anisotropy. This is illustrated by the two plots on the right in Figure~
\ref{fig: anisotropic distribution on S2}, for different times.


\section*{Code}
The code used for the experiments is available in the Theano Geometry software package
\footnote{\url{http://bitbucket.org/stefansommer/theanogeometry}}.
The implementation uses automatic differentiation libraries extensively for the geometry computations as is further described in \cite{kuhnel_differential_2019}.

\section*{Acknowledgement} 
The work presented is supported by the CSGB Centre for Stochastic Geometry and Advanced Bioimaging funded by a grant from the Villum foundation, the Villum Foundation grants 22924 and 40582, and the Novo Nordisk Foundation grant NNF18OC0052000.

\bibliography{bibfile,library}

\begin{thebibliography}{999}

\bibitem[Pedersen(1995)]{pedersen1995consistency}
Pedersen, A.R.
\newblock Consistency and asymptotic normality of an approximate maximum
  likelihood estimator for discretely observed diffusion processes.
\newblock {\em Bernoulli} {\bf 1995}, pp. 257--279.

\bibitem[Bladt \em{et~al.}(2014)Bladt, S{\o}rensen, et~al.]{bladt2014simple}
Bladt, M.; S{\o}rensen, M.;  et~al.
\newblock Simple simulation of diffusion bridges with application to likelihood
  inference for diffusions.
\newblock {\em Bernoulli} {\bf 2014}, {\em 20},~645--675.

\bibitem[Bladt \em{et~al.}(2016)Bladt, Finch, and
  S{\o}rensen]{bladt2016simulation}
Bladt, M.; Finch, S.; S{\o}rensen, M.
\newblock Simulation of multivariate diffusion bridges.
\newblock {\em Journal of the Royal Statistical Society: Series B: Statistical
  Methodology} {\bf 2016}, pp. 343--369.

\bibitem[Bui \em{et~al.}(2021)Bui, Pokern, and Dellaportas]{bui2021inference}
Bui, M.N.; Pokern, Y.; Dellaportas, P.
\newblock Inference for partially observed Riemannian Ornstein--Uhlenbeck
  diffusions of covariance matrices.
\newblock {\em arXiv preprint arXiv:2104.03193} {\bf 2021}.

\bibitem[Delyon and Hu(2006)]{delyon_simulation_2006}
Delyon, B.; Hu, Y.
\newblock Simulation of Conditioned Diffusion and Application to Parameter
  Estimation.
\newblock {\em Stochastic Processes and their Applications} {\bf 2006}, {\em
  116},~1660--1675.

\bibitem[Jensen \em{et~al.}(2019)Jensen, Mallasto, and
  Sommer]{jensen2019simulation}
Jensen, M.H.; Mallasto, A.; Sommer, S.
\newblock Simulation of Conditioned Diffusions on the Flat Torus.
\newblock  International Conference on Geometric Science of Information.
  Springer,  2019, pp. 685--694.

\bibitem[Jensen and Sommer(2021)]{jensen_simulation_2021}
Jensen, M.H.; Sommer, S.
\newblock Simulation of {{Conditioned Semimartingales}} on {{Riemannian
  Manifolds}}.
\newblock {\em arXiv:2105.13190} {\bf 2021},
  \href{http://xxx.lanl.gov/abs/2105.13190}{{\normalfont [2105.13190]}}.

\bibitem[van~der Meulen and Schauer(2017)]{van2017bayesian}
van~der Meulen, F.; Schauer, M.
\newblock Bayesian estimation of discretely observed multi-dimensional
  diffusion processes using guided proposals.
\newblock {\em Electronic Journal of Statistics} {\bf 2017}, {\em
  11},~2358--2396.

\bibitem[Papaspiliopoulos and Roberts(2012)]{papaspiliopoulos2012importance}
Papaspiliopoulos, O.; Roberts, G.
\newblock Importance sampling techniques for estimation of diffusion models.
\newblock {\em Statistical methods for stochastic differential equations} {\bf
  2012}, pp. 311--340.

\bibitem[Schauer \em{et~al.}(2017)Schauer, Van Der~Meulen, Van~Zanten,
  et~al.]{schauer2017guided}
Schauer, M.; Van Der~Meulen, F.; Van~Zanten, H.;  et~al.
\newblock Guided proposals for simulating multi-dimensional diffusion bridges.
\newblock {\em Bernoulli} {\bf 2017}, {\em 23},~2917--2950.

\bibitem[Sommer \em{et~al.}(2017)Sommer, Arnaudon, Kuhnel, and
  Joshi]{sommer_bridge_2017}
Sommer, S.; Arnaudon, A.; Kuhnel, L.; Joshi, S.
\newblock Bridge {{Simulation}} and {{Metric Estimation}} on {{Landmark
  Manifolds}}.
\newblock  Graphs in {{Biomedical Image Analysis}}, {{Computational Anatomy}}
  and {{Imaging Genetics}}. {Springer},  2017, Lecture {{Notes}} in {{Computer
  Science}}, pp. 79--91.

\bibitem[Bui(2022)]{bui2022inference}
Bui, M.N.
\newblock Inference on Riemannian Manifolds: Regression and Stochastic
  Differential Equations.
\newblock PhD thesis, UCL (University College London),  2022.

\bibitem[Fisher(1953)]{fisher_dispersion_1953}
Fisher, R.
\newblock Dispersion on a {{Sphere}}.
\newblock {\em Proceedings of the Royal Society of London A: Mathematical,
  Physical and Engineering Sciences} {\bf 1953}, {\em 217},~295--305.
\newblock
  doi:{\changeurlcolor{black}\href{https://doi.org/10.1098/rspa.1953.0064}{\detokenize{10.1098/rspa.1953.0064}}}.

\bibitem[Kent(1982)]{kent_fisher-bingham_1982}
Kent, J.T.
\newblock The {{Fisher-Bingham Distribution}} on the {{Sphere}}.
\newblock {\em Journal of the Royal Statistical Society. Series B
  (Methodological)} {\bf 1982}, {\em 44},~71--80.

\bibitem[Thompson(2018)]{thompson2018brownian}
Thompson, J.
\newblock Brownian bridges to submanifolds.
\newblock {\em Potential Analysis} {\bf 2018}, {\em 49},~555--581.

\bibitem[{Garc{\'i}a-Portugu{\'e}s}
  \em{et~al.}(2017){Garc{\'i}a-Portugu{\'e}s}, S{\o}rensen, Mardia, and
  Hamelryck]{garcia-portugues_langevin_2017}
{Garc{\'i}a-Portugu{\'e}s}, E.; S{\o}rensen, M.; Mardia, K.V.; Hamelryck, T.
\newblock Langevin Diffusions on the Torus: Estimation and Applications.
\newblock {\em Statistics and Computing} {\bf 2017}, pp. 1--22.
\newblock
  doi:{\changeurlcolor{black}\href{https://doi.org/10.1007/s11222-017-9790-2}{\detokenize{10.1007/s11222-017-9790-2}}}.

\bibitem[Hamelryck \em{et~al.}(2006)Hamelryck, Kent, and
  Krogh]{hamelryck_sampling_2006}
Hamelryck, T.; Kent, J.T.; Krogh, A.
\newblock Sampling {{Realistic Protein Conformations Using Local Structural
  Bias}}.
\newblock {\em PLoS Computational Biology} {\bf 2006}, {\em 2}.
\newblock
  doi:{\changeurlcolor{black}\href{https://doi.org/10.1371/journal.pcbi.0020131}{\detokenize{10.1371/journal.pcbi.0020131}}}.

\bibitem[Pennec \em{et~al.}(2006)Pennec, Fillard, and
  Ayache]{pennec_riemannian_2006}
Pennec, X.; Fillard, P.; Ayache, N.
\newblock A {{Riemannian Framework}} for {{Tensor Computing}}.
\newblock {\em Int. J. Comput. Vision} {\bf 2006}, {\em 66},~41--66.

\bibitem[Vaillant \em{et~al.}(2004)Vaillant, Miller, Younes, and
  Trouv{\'e}]{vaillant_statistics_2004}
Vaillant, M.; Miller, M.; Younes, L.; Trouv{\'e}, A.
\newblock Statistics on Diffeomorphisms via Tangent Space Representations.
\newblock {\em NeuroImage} {\bf 2004}, {\em 23},~S161--S169.
\newblock
  doi:{\changeurlcolor{black}\href{https://doi.org/10.1016/j.neuroimage.2004.07.023}{\detokenize{10.1016/j.neuroimage.2004.07.023}}}.

\bibitem[Yang(2011)]{yang_means_2011}
Yang, L.
\newblock Means of Probability Measures in {{Riemannian}} Manifolds and
  Applications to Radar Target Detection.
\newblock PhD thesis, Poitiers University,  2011.

\bibitem[Grenander(1963)]{grenander_probabilities_1963}
Grenander, U.
\newblock {\em Probabilities on {{Algebraic Structures}}}; {Wiley}: {New York
  and London},  1963.

\bibitem[Pennec \em{et~al.}(2020)Pennec, Sommer, and
  Fletcher]{pennec_riemannian_2020}
Pennec, X.; Sommer, S.; Fletcher, T.
\newblock {\em Riemannian {{Geometric Statistics}} in {{Medical Image
  Analysis}}}; {Elsevier},  2020.

\bibitem[Liao(2004)]{liao2004levy}
Liao, M.
\newblock {\em L{\'e}vy processes in Lie groups}; Vol. 162, Cambridge
  university press,  2004.

\bibitem[Shigekawa(1984)]{shigekawa1984transformations}
Shigekawa, I.
\newblock Transformations of the Brownian motion on a Riemannian symmetric
  space.
\newblock {\em Zeitschrift f{\"u}r Wahrscheinlichkeitstheorie und Verwandte
  Gebiete} {\bf 1984}, pp. 493--522.

\bibitem[Jensen and Sommer(2022)]{jensen2022mean}
Jensen, M.H.; Sommer, S.
\newblock Mean Estimation on the Diagonal of Product Manifolds.
\newblock {\em Algorithms} {\bf 2022}, {\em 15},~92.

\bibitem[Pennec \em{et~al.}(2019)Pennec, Sommer, and
  Fletcher]{pennec2019riemannian}
Pennec, X.; Sommer, S.; Fletcher, T.
\newblock {\em Riemannian Geometric Statistics in Medical Image Analysis};
  Academic Press,  2019.

\bibitem[Kendall(1987)]{kendall1987radial}
Kendall, W.S.
\newblock The radial part of Brownian motion on a manifold: a semimartingale
  property.
\newblock {\em The Annals of Probability} {\bf 1987}, {\em 15},~1491--1500.

\bibitem[Barden and Le(1997)]{barden1997some}
Barden, D.; Le, H.
\newblock Some consequences of the nature of the distance function on the cut
  locus in a Riemannian manifold.
\newblock {\em Journal of the LMS} {\bf 1997}, pp. 369--383.

\bibitem[Le and Barden(1995)]{le1995ito}
Le, H.; Barden, D.
\newblock It{\^o} correction terms for the radial parts of semimartingales on
  manifolds.
\newblock {\em Probability theory and related fields} {\bf 1995}, {\em
  101},~133--146.

\bibitem[Hsu(2002)]{hsu2002stochastic}
Hsu, E.P.
\newblock {\em Stochastic analysis on manifolds}; Vol.~38, AMS,  2002.

\bibitem[Thompson(2015)]{thompson2015submanifold}
Thompson, J.
\newblock Submanifold bridge processes.
\newblock PhD thesis, University of Warwick,  2015.

\bibitem[Thompson and Li(2015)]{thompson_submanifold_2015}
Thompson, J.; Li, X.M.
\newblock Submanifold Bridge Processes.
\newblock PhD thesis, University of Warwick, {Coventry},  2015.

\bibitem[van~der Meulen and Schauer(2018)]{van2018bayesian}
van~der Meulen, F.; Schauer, M.
\newblock Bayesian estimation of incompletely observed diffusions.
\newblock {\em Stochastics} {\bf 2018}, {\em 90},~641--662.

\bibitem[Mider \em{et~al.}(2021)Mider, Schauer, and van~der
  Meulen]{mider2021continuous}
Mider, M.; Schauer, M.; van~der Meulen, F.
\newblock Continuous-discrete smoothing of diffusions.
\newblock {\em Electronic Journal of Statistics} {\bf 2021}, {\em
  15},~4295--4342.

\bibitem[Arnaudon \em{et~al.}(2022)Arnaudon, van~der Meulen, Schauer, and
  Sommer]{arnaudon2022diffusion}
Arnaudon, A.; van~der Meulen, F.; Schauer, M.; Sommer, S.
\newblock Diffusion bridges for stochastic Hamiltonian systems and shape
  evolutions.
\newblock {\em SIAM Journal on Imaging Sciences} {\bf 2022}, {\em
  15},~293--323.

\bibitem[Hansen \em{et~al.}(2021{\natexlab{a}})Hansen, Eltzner, and
  Sommer]{hansen2021diffusion}
Hansen, P.; Eltzner, B.; Sommer, S.
\newblock Diffusion Means and Heat Kernel on Manifolds.
\newblock {\em arXiv preprint arXiv:2103.00588} {\bf 2021}.

\bibitem[Hansen \em{et~al.}(2021{\natexlab{b}})Hansen, Eltzner, Huckemann, and
  Sommer]{hansen2021diffusiongeometric}
Hansen, P.; Eltzner, B.; Huckemann, S.F.; Sommer, S.
\newblock Diffusion Means in Geometric Spaces.
\newblock {\em arXiv preprint arXiv:2105.12061} {\bf 2021}.

\bibitem[K{\"u}hnel \em{et~al.}(2019)K{\"u}hnel, Sommer, and
  Arnaudon]{kuhnel_differential_2019}
K{\"u}hnel, L.; Sommer, S.; Arnaudon, A.
\newblock Differential Geometry and Stochastic Dynamics with Deep Learning
  Numerics.
\newblock {\em Applied Mathematics and Computation} {\bf 2019}, {\em
  356},~411--437.
\newblock
  doi:{\changeurlcolor{black}\href{https://doi.org/10.1016/j.amc.2019.03.044}{\detokenize{10.1016/j.amc.2019.03.044}}}.

\end{thebibliography}

\end{document}